\documentclass[useAMS,usenatbib,usegraphicx]{mn2e}

\newcommand{\der}{\mathrm{d}}

\title[Regularized orbit models for the elliptical NGC 4807 and its dark halo]{Regularized orbit models unveiling the stellar structure and dark matter halo of the Coma elliptical NGC 4807}
\author[J. Thomas et al.]{J. Thomas$^{1,2}$\thanks{E-mail: jthomas@mpe.mpg.de}, R. P. Saglia$^{2}$, R. Bender$^{1,2}$, 
D. Thomas$^{2}$, K. Gebhardt$^{3}$, 
\newauthor J. Magorrian$^{4}$, E.~M. Corsini$^{5}$, G. Wegner$^{6}$\\
$^{1}$Universit\"atssternwarte M\"unchen, Scheinerstra\ss e 1, D-81679 M\"unchen, Germany\\
$^{2}$Max-Planck-Institut f\"ur Extraterrestrische Physik, Giessenbachstra\ss e, D-85748 Garching, Germany\\
$^{3}$Department of Astronomy, University of Texas at Austin, C1400, Austin, TX78712, USA\\
$^{4}$Theoretical Physics, Department of Physics, University of Oxford, 1 Keble Road, Oxford U.K., OX1 3NP\\
$^{5}$Dipartimento di Astronomia, Universit\`a di Padova, vicolo dell'Osservatorio 2, I-35122 Padova, Italy\\
$^{6}$Department of Physics and Astronomy, 6127 Wilder Laboratory, Dartmouth College, Hanover, NH 03755-3528, USA}
\begin{document}

\date{Accepted 1988 December 15. Received 1988 December 14; in original form 1988 October 11}

\pagerange{\pageref{firstpage}--\pageref{lastpage}} \pubyear{2002}

\maketitle

\label{firstpage}

\begin{abstract}
This is the second in a series of papers dedicated to unveil the mass structure and orbital
content of a sample of flattened early-type galaxies in the Coma cluster. The ability of
our orbit libraries to reconstruct internal stellar motions and the mass 
composition of a typical elliptical in the sample is investigated by means of Monte-Carlo 
simulations of isotropic rotator models. The simulations allow a determination of the optimal 
amount of regularization needed in the orbit superpositions. It is shown that under 
realistic observational conditions
and with the appropriate regularization internal velocity moments can be reconstructed to
an accuracy of $\approx 15$ per cent; the same accuracy can be achieved for the circular
velocity and dark matter fraction. In contrast, the flattening of the halo remains
unconstrained. Regularized orbit superpositions are applied to 
a first galaxy in our sample, NGC 4807, for which stellar kinematical
observations extend to $3 \, r_\mathrm{eff}$. The galaxy seems dark matter dominated 
outside $r> 2 \, r_\mathrm{eff}$. Logarithmic dark matter potentials are consistent with
the data, as well as NFW-profiles (Navarro, Frenk \& White 1996), mimicking logarithmic 
potentials over the observationally sampled radial range. In both cases, the derived
stellar mass-to-light ratio $\Upsilon$ agrees well with independently obtained
mass-to-light ratios from stellar population analysis. The achieved accuracy is
$\Delta \Upsilon \approx 0.5$. Kinematically, NGC 4807 is
characterized by mild radial anisotropy outside $r> 0.5 \, r_\mathrm{eff}$, becoming
isotropic towards the center. Our orbit models
hint at either a distinct stellar component or weak triaxiality in the outer
parts of the galaxy.
\end{abstract}



\section{Introduction}
Interpreting the stellar kinematical data of ellipticals in terms of galaxy structural
parameters requires knowing the gravitational potential as well as the 
distribution of stellar orbits, which -- due to projection
effects -- is not directly given by observations. In equilibrium stellar systems, the
phase-space distribution function (DF) describing the orbital state depends
on phase-space coordinates only through the integrals of motion, admitted by the actual
potential (Jeans theorem; e.g. \citealt{Bin87}). 

In cases where integrals of motion can be expressed (or approximated) in terms
of elementary functions, the DF can be parameterized explicitly. Several round as well as a 
couple of flattened ellipticals have been modelled based on this approach
(e.g. \citealt{DehGer94}; \citealt{Car95}; \citealt{Qia95}; \citealt{Dej96}; \citealt{G98}; 
\citealt{Ems99}; \citealt{Mat99}; \citealt{Sta99}; \citealt{Kr00}; \citealt{Sag00}; 
\citealt{G01}). On the other hand, \citet{S79} orbit superposition modelling technique
provides fully general dynamical models for {\it any} axisymmetric or triaxial
potential and has been successfully applied to a growing number of early-type galaxies
(e.g. \citealt{R97}; \citealt{Cre98}; \citealt{Cre00}; \citealt{Geb00}; 
\citealt{Cap02}; \citealt{Ver02}; \citealt{Geb03}; \citealt{R03}; \citealt{Cop04}; 
\citealt{Val04}; \citealt{Kra05}).

Even if the number of dynamical models is steadily increasing, 
the only comprehensive investigation of elliptical galaxy DFs 
addressing the question of dark matter in these systems is still the
basis-function based spherical modeling of 21 round galaxies by 
\citet{Kr00} and \citet{G01}. To extend the results found there and in the handful of studies
of individual objects quoted above, we started a project
aimed to probe a sample of {\it flattened} early-type galaxies in the Coma cluster,
collecting major \citep{coma1} and minor axis \citep{coma2} kinematical data. 
The goal is to investigate the dynamical structure and dark matter
content of these galaxies.  

Most present-day orbit superpositions, conceptionally based on
Schwarzschild's original implementation, do not 
automatically provide the entire phase-space DF, but only orbital occupation numbers or 
weights, respectively, characterizing the total amount of light carried by each orbit.
In principle, changing the orbit sampling strategy allows one to infer the DF from any
orbit superposition \citep{Haf00}, but this approach has not yet been followed fully.
To take advantage of both, the full generality of orbit superpositions and the availability 
of DFs, we extended the orbit superposition code of 
\citet{maxspaper} and \citet{Geb00} to reconstruct phase-space DFs
in the axisymmetric case \citep{comadyn1}.

Recovering stellar DFs from photometric and kinematic observations -- in particular the
application of non-parametric methods like orbit superpositions -- invokes 
regularization in order to pick up smooth phase-space distributions (e.g. 
\citealt{maxspaper}; \citealt{Mer93}). 
Different regularization schemes have been applied in the context of 
orbit libraries, among them variants of minimizing occupation number gradients in orbit 
space \citep{Z96} and maximum entropy \citep{maxspaper}. The proper amount of regularization 
has thereby commonly been adjusted to the data (e.g. \citealt{R97}; \citealt{Ver02}; 
Richstone et al., in preparation). 

One aim of this paper is to readdress the question of how much regularization is
needed to recover galaxy internal structural properties from observations with
spatial sampling and noise typical for our sample of Coma galaxies. To this end we 
study observationally motivated reference models under realistic observational 
conditions and optimize regularization with respect to the reconstruction of intrinsic
input-model properties (see e.g. \citealt{G98}; \citealt{Cre99}).
By simulating and recovering reference galaxies we also determine to
which degree the internal mass structure and orbital content are constrained by
observational data in our sample. Furthermore $\chi^2$-statistics are measured to 
assign confidence intervals to real galaxy orbit superpositions.

To demonstrate the prospects of regularized orbit superpositions for our project 
we also present an application to one galaxy. We have chosen the faint
giant E2 elliptical NGC 4807 ($M_B = -20.76$ for $H_0 = 69 \, \mathrm{km/s/Mpc}$ 
from Hyperleda) for the following reasons:
(1) The galaxy has a prominent boxy photometric feature in the outer parts constraining 
its inclination. (2) Stellar kinematic data reach out to 3 $r_\mathrm{eff}$ 
along the major axis probing its dark halo outside the galaxy main-body. (3) NGC 4807 lacks
significant 
minor-axis rotation and isophotal twist and thus is an ideal target for axisymmetric
modelling. (4) The galaxy is only mildly flattened and the dynamical models can be compared
with earlier studies of similar galaxies done mostly in the spherical approximation.

The paper is organized as follows. Sec.~\ref{data} summarizes the observations of NGC 4807, 
Sec.~\ref{library} outlines the orbit superposition technique. In Sec.~\ref{alpha} 
Monte-Carlo simulations performed to derive the optimal regularization are described and
the achieved accuracy of galaxy reconstructions follows in 
Sec.~\ref{confidence}. Orbit models 
for NGC 4807 are presented in Secs.~\ref{mass5975}-\ref{phasstruc}. 
The paper is closed with a combined summary and discussion of the results in 
Sec.~\ref{discussion}.


\section[]{NGC 4807: model input}
\label{data}
In the following we assume a distance $d = 100 \, \mathrm{Mpc}$ to NGC 4807
(corresponding to $H_0 = 69 \, \mathrm{km/s/Mpc}$) and take $r_\mathrm{eff} = 6.7$ 
arcsec as its effective radius \citep{coma1}.

\subsection{Photometric data}
\label{photometry}
The photometric data are combined from two different sources. For the outer parts of NGC 4807 
Kron-Cousins $R_C$ band CCD photometry is drawn from \citet{coma1}, consisting of
profiles for the surface brightness $\mu_R$, ellipticity $\epsilon$ and
isophotal shape parameters $a_4$ and $a_6$ out to $\approx 5.5 \, r_\mathrm{eff}$
(see \citealt{bm87} for a definition of $a_n$).
A seeing of 2 arcsec during the observations causes the profiles to be unresolved in 
their central parts (at $d=100 \, \mathrm{Mpc}$ one arcsec corresponds to 
$0.485 \, \mathrm{kpc}$). 
To increase the central resolution, the ground-based data are supplemented by corresponding
profiles for $\mu_V,\epsilon,a_4$ and $a_6$ extracted from archival HST V band data 
(Principal Investigator: John Lucey; Proposal ID: 5997).

The two surface brightness profiles $\mu_R$ and $\mu_V$ 
are joined by shifting the HST V band according to 
the average $\langle \mu_R-\mu_V \rangle$ 
taken over the region $0.75 \, r_\mathrm{eff} \le R \le 3 \, r_\mathrm{eff}$, 
where both data sets overlap and seeing effects are negligible. The shift 
$\langle \mu_R-\mu_V \rangle$ is well defined with a rms of only $0.015 \, \mathrm{mag}$. 

For the orbit models ground-based photometry is used outside $R \ge 3 \, r_\mathrm{eff}$ 
and HST data inside $R < 3 \, r_\mathrm{eff}$. Fig.~\ref{depro} shows the 
combined $\mu_R$, $\epsilon$, $a_4$ and $a_6$ profiles applied for the modelling.

\subsection{Deprojection}
\label{deprojection}

\begin{figure}
\includegraphics[width=84mm,angle=0]{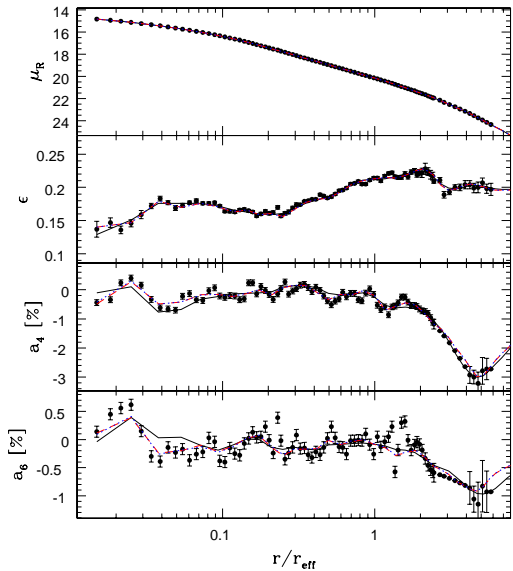}
\caption{Photometric data for NGC 4807 (dots with error bars, from top to bottom: 
surface brightness $\mu_R$, ellipticity $\epsilon$, isophotal shape parameters 
$a_4$ and $a_6$) and three deprojections:
edge-on deprojection (solid); $i=50\degr$-deprojection (dashed); disky $i=50\degr$-
deprojection (dotted).}
\label{depro}
\end{figure}

In the implementation of Schwarzschild's modelling technique used here, orbit models are not
directly fitted to observed photometry. Instead, the deprojected luminosity density 
is used as a boundary condition for any orbit superposition (see Sec.~\ref{fit}).

Deprojection of axisymmetric galaxies is unique only for edge-on systems
(inclination $i=90\degr$; \citealt{ryb87}). For any inclination $i<90\degr$ disk-like
konus-densities can be added to the luminosity profile without affecting its projected
appearance for any $i^\prime < i$ \citep{GerBi96}. From $i=90\degr$ to $i=0\degr$ the 
variety of different luminosity models projecting to the same galaxy image generally 
increases.

Deprojections for NGC 4807 are obtained with the program of \citet{mag99}. The code 
allows one to explore the full range of luminosity densities consistent
with the photometric data by forcing the deprojection towards different internal
shapes. For example, at any given inclination $i$ the goodness of fit
of the deprojection can be 
penalized towards any degree of internal boxiness and diskiness, respectively. The 
deprojections are fitted without seeing-correction, since our joint photometry includes 
ground-based data only outside $R \ge 3 \, r_\mathrm{eff}$, where seeing effects are 
negligible. Outside the last measured photometric data point ($R \ge 5.5 \, r_\mathrm{eff}$), 
the photometry is extrapolated by a de Vaucouleurs profile fitted to the inner parts of the
galaxy. The isophotes outside $R > 5.5 \, R_\mathrm{eff}$ are assumed to be
perfect ellipses with a constant flattening corresponding to the galaxy's outermost measured 
ellipticity. 

\begin{figure}
\includegraphics[width=84mm,angle=0]{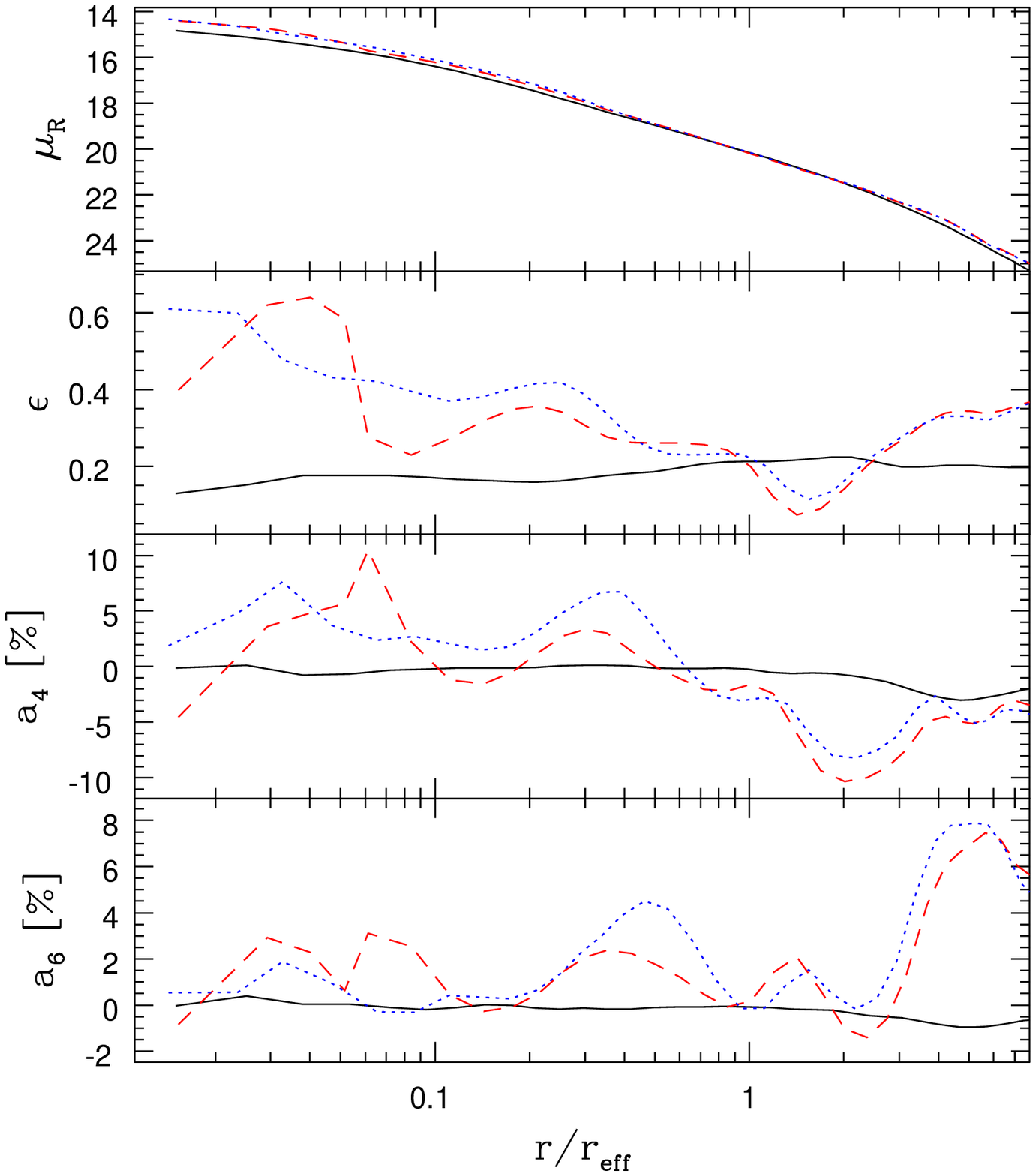}
\includegraphics[width=84mm,angle=0]{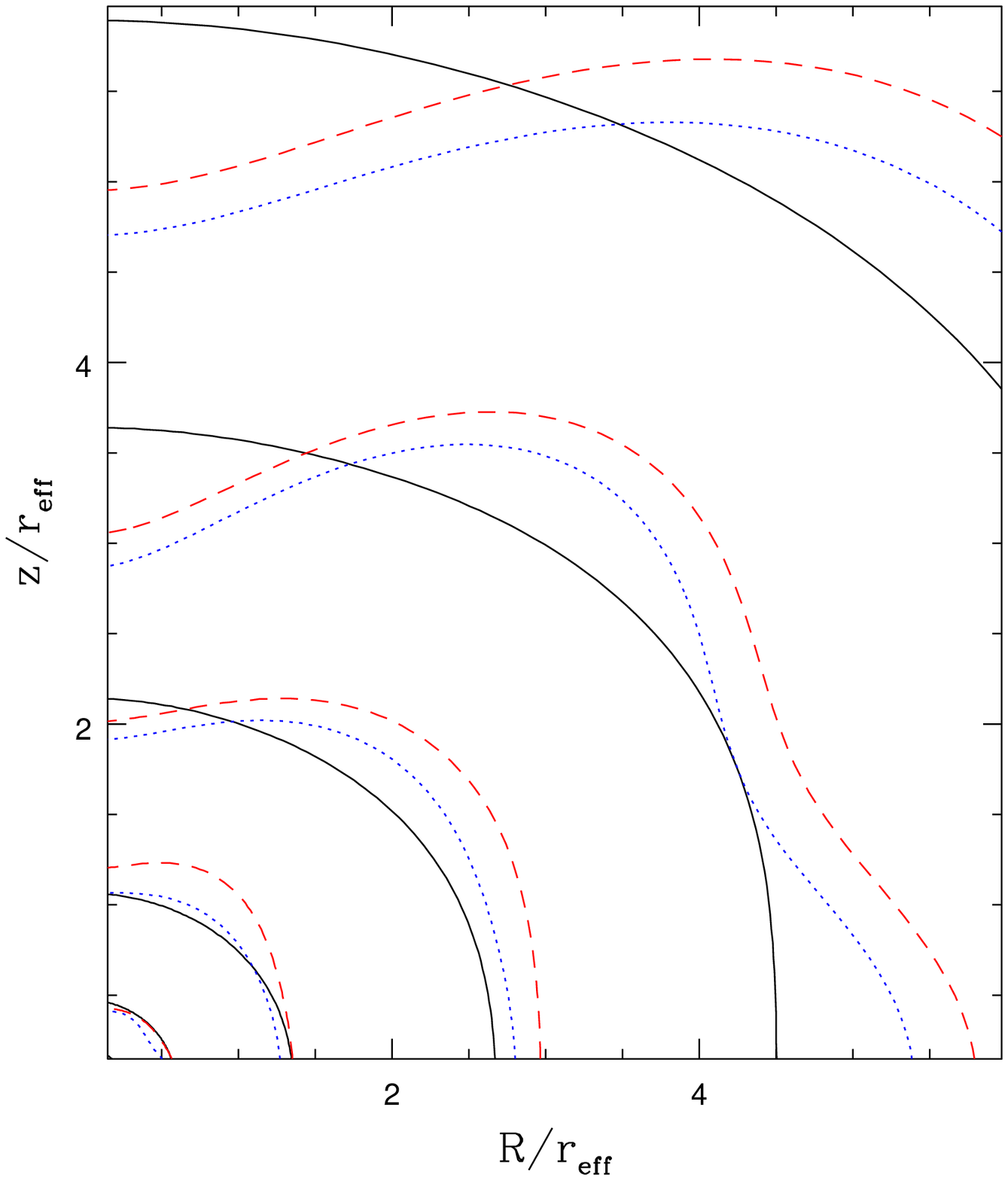}
\caption{Projected appearance of the three deprojections shown in Fig.~\ref{depro} when
seen edge-on. The top panel shows isophotal shape parameters (cf. 
Fig.~\ref{depro}), the bottom panel displays the corresponding isophotes. Solid:
$i=90\degr$ deprojection; dashed: $i=50\degr$-deprojection; 
 dotted: disky $i=50\degr$-deprojection.}
\label{depro_rpi90}
\end{figure}

From the average flattening $\langle q \rangle = 0.8$ of NGC 4807 we expect an 
inclination angle $i > 38\degr$. Lower viewing angles would require density distributions
intrinsically flatter than E7, which are not observed. In Fig.~\ref{depro} three 
representative deprojections are overplotted to the photometric data: the (unique) 
$i=90\degr$-deprojection; a deprojection at $i=50\degr$ without any shape penalty; a disky 
$i=50\degr$-deprojection. All three luminosity models are equally consistent with the data. 
The differences between the models are illustrated in Fig.~\ref{depro_rpi90}, which displays 
the isophotal shape parameters and the corresponding isophotes of the deprojections as they 
appear viewed from edge-on. At this viewing angle internal density distortions show up the 
strongest.

As Fig.~\ref{depro_rpi90} reveals, both deprojections at $i=50\degr$ are heavily boxy around 
$R \approx 2 \, r_\mathrm{eff}$ ($a_4 \approx -10$ per cent for the deprojection without 
shape preference and $a_4 \approx -8$ per cent for the disky deprojection). These distortions
are a reflection of the drop in $a_4$ at $R > 3\, r_\mathrm{eff}$ in the data of NGC 4807.
Projected density distortions progressively strengthen in deprojection from $i=90\degr$ to 
$i=0\degr$ and assuming $i=50\degr$ already causes a considerable amplification. Near the 
center -- where the observed isophotes are consistent with being purely elliptical -- 
the disky deprojection appears smoother with on average larger $a_4$, albeit lacking the
strong $a_4$-peak occurring in the deprojection without shape preference at $R=0.05 \,
r_\mathrm{eff}$.

We have constructed orbit models for both, the $i=90\degr$ and the non-disky $i=50\degr$ 
deprojections, but the heavily distorted density distributions at $i=50\degr$ lead us to 
expect that we likely view NGC 4807 close to edge-on.

\subsection{Kinematic data}
\label{kinematics}
The kinematic data are described in \citet{coma1} and \citet{coma2}. They consist
of two long-slit spectra for the major and the minor axis, respectively. Profiles of 
lower order Gauss-Hermite coefficients $\gamma_0$, $v$, $\sigma$, $H_3$ and $H_4$ 
\citep{Ger93,vdMF93} parameterizing the line-of-sight velocity distribution (LOSVD) reach 
out to $3 \, r_\mathrm{eff}$ along the major-axis and out to $0.6 \, r_\mathrm{eff}$ along 
the minor-axis. Our orbit models are not directly fitted to the observed Gauss-Hermite 
parameters, but to binned LOSVDs. Therefore the Gauss-Hermite parameters are transformed into 
suitable LOSVDs as follows. First, the Gauss-Hermite series according to the observed 
$\gamma_0$, $v$, $\sigma$, $H_3$ and $H_4$ is evaluated at 1000 values of 
line-of-sight velocity $v_\mathrm{los}$, linearly spaced in the range 
$-4 \, \sigma \le v_\mathrm{los} \le 4 \, \sigma$. At each of the sampled projected 
velocities error bars $\Delta \mathrm{LOSVD}$ 
are assigned to the LOSVD from the errors in the corresponding
Gauss-Hermite parameters by means of Monte-Carlo simulations. The LOSVD is then binned 
appropriately for comparison with the orbit library (see Sec.~\ref{grids} below).

For NGC 4807, the scatter in kinematical data from different sides of the galaxy is mostly 
negligible compared with the statistical errors in the data. Therefore, the orbit models
are fitted to a symmetrized data set, derived by averaging the Gauss-Hermite parameters
from each side. The largest scatter appears in the outermost major-axis $H_4$-measurement. 
To infer the impact of its uncertainty on the reconstructed DF, we separately refitted the 
mass model that most closely matches the symmetrized data set, to the kinematics of the 
two different sides of NGC 4807. Both fits lead to similar models in terms of DF and internal
kinematical structure, suggesting that the asymmetry in this single measurement does
not largely affect our results.


\section[]{Orbit superposition models}
\label{library}
Our method of setting up orbit libraries to construct best fitting models is described in
detail in Richstone et al. (in preparation) and in \citet{comadyn1}. Here, we only briefly
outline some aspects of the method relevant to the present paper.

To recover the mass distribution of a given galaxy by orbit libraries, a grid of 
(parameterized) trial potentials is probed. In 
each trial potential (Sec.~\ref{masscomp}) a large set of orbits 
is calculated (Sec.~\ref{orbits}) and an orbit superposition is constructed that best matches 
the observational constraints (Sec.~\ref{fit}). The best-fitting mass parameters with
corresponding errors then follow from a $\chi^2$-analysis.

\subsection{Basic grids}
\label{grids}
For comparison with observations the meridional plane of the orbit model as well as
its projection onto the plane of the sky are divided into bins in radius 
$r$ and polar angle $\vartheta$. The
grid in radius is logarithmic in the outer parts and becomes linear in the inner parts; the
angular bins cover equal steps in $\sin \vartheta$ (see Richstone et al,
in preparation, for details). For NGC 4807 the models are calculated in 
$N_r \times N_\vartheta = 200$ bins with $N_r=20$ and $N_\vartheta=10$. Each spatial bin 
of the model's sky-projection is subdivided into 
$N_\mathrm{vel} = 15$ bins in projected velocity 
for the purpose of fitting the library to the observed kinematics.

\subsection{Luminous and dark mass distributions}
\label{masscomp}
\paragraph*{Luminous matter.}
We assume that the luminous mass of NGC 4807 is sufficiently characterized
by a constant stellar mass-to-light ratio $\Upsilon$ (see also Sec.~\ref{discussion} and
Fig.~\ref{projml} for a verification of this assumption). The stellar mass density then reads
$\Upsilon \, \nu$, where the luminosity 
density $\nu$ is taken from the deprojections of Sec.~\ref{deprojection}.

\paragraph*{Dark matter.}
Modelling of spiral and dwarf galaxy rotation curves indicates shallow inner dark matter 
density distributions, i.e. logarithmic density slopes 
$\gamma \equiv \der \ln(\rho_\mathrm{DM}) / \der \ln(r) \approx 0$ (e.g. \citealt{deB01}; 
\citealt{G04}). Flat dark matter density cores also fit to the kinematics 
of round ellipticals \citep{G01}, although \citet{R97} found steeper profiles consistent with 
observations of NGC 2434. Some optical or radio rotation curves also allow density slopes 
up to $\gamma \leq -1$, but they do not prefer steep profiles \citep{S03}.

Shallow dark matter density profiles conflict with predictions of (pure dark matter) 
cosmological $N$-body simulations. In the $\Lambda$CDM scenario, dark matter 
distributions with central density cusps around $-1.5 < \gamma < -1$ are found
(e.g. \citealt{Mor99}; \citealt{N04}). Similar steep profiles emerge in
warm dark matter (\citealt{C00}; \citealt{K02}), whereas self-interacting dark matter 
offers central $\gamma \approx 0$ \citep{D01}. 

To probe the whole diversity of theoretically and observationally motivated profiles
we allow for two different dark matter distributions, 
one with central $\gamma = 0$, representative of shallow profiles and another with
central $\gamma = -1$, representative of the steeper cases.

\subparagraph*{Cored profiles.}
A dark matter distribution that provides asymptotically 
flat circular velocity curves in combination with flat ($\gamma = 0$) inner density cores
is given by the logarithmic potential \citep{Bin87}
\begin{equation}
\label{nispot}
\Phi_\mathrm{LOG}(R,z) = \frac{v_c^2}{2} \ln \left( 
r_c ^2 + R^2 + \frac{z^2}{q^2} \right),
\end{equation}
where $R=r\cos(\vartheta)$, $z=r\sin(\vartheta)$ are cartesian coordinates in the 
meridional plane and $\vartheta = 0\degr$ corresponds to the equator.
The density distribution generating $\Phi_\mathrm{LOG}$ reads \citep{Bin87}
\begin{equation}
\label{nis}
\rho_\mathrm{LOG}(R,z) \propto v_c^2
\frac{(2q^2+1)r_c^2+R^2+2(1-\frac{1}{2q^2})z^2}
{(r_c^2+R^2+z^2/q^2)^2\;q^2}
\end{equation}
and is positive everywhere for $q \in [1/\sqrt{2},1]$. The flattening of the density 
distribution (\ref{nis}) differs from that of the potential $q$. It is generally smaller 
and varies with radius. 
In the following, we will only consider cored profiles with $q=1.0$ (spherical). 
Together with the assumption of a constant
mass-to-light ratio $\Upsilon$ models with dark matter distributions according to
equation (\ref{nis}) can be seen as analogues to maximum-disk models of spiral 
galaxies and have been called maximum-stellar-mass models \citep{G98}.

\subparagraph*{Cuspy profiles.}
A representative cuspy mass distribution fitting simulated dark matter halos over a 
wide range of radii is the Navarro-Frenk-White (NFW) profile \citep{nfw96}
\begin{equation}
\label{nfw}
\rho_\mathrm{NFW}(r,r_s,c) \propto \frac{\delta_c}{(r/r_s)(1+r/r_s)^2},
\end{equation}
with a central logarithmic density slope $\gamma = -1$. The parameter $\delta_c$
in equation (\ref{nfw}) is related to a concentration parameter $c$ via
\begin{equation}
\delta_c = \frac{200}{3} \frac{c^3}{\mathrm{ln}(1+c) - c/(1+c)}.
\end{equation}
By the substitution
$r \rightarrow r \sqrt{\cos^2 (\vartheta) + \sin^2 (\vartheta) / q^2}$ equation (\ref{nfw})
provides halos with a constant flattening $q$.

In CDM cosmology
the two parameters $c$ and $r_s$ turn out to be correlated in the 
sense that higher mass halos are less concentrated, with some scatter due to different 
mass assembly histories (\citealt{nfw96}; \citealt{J00}; \citealt{W02}).
The corresponding relation reads
\begin{equation}
\label{family}
r_s^3 \propto 10^{(A-\log c)/B} \left( 200 \, \frac{4 \pi}{3} \, c^3 \right)^{-1}.
\end{equation}
Here, we take $A=1.05$ and $B=0.15$ (\citealt{nfw96}; \citealt{R97}), which -- for
the concentrations $5<c<25$ considered here -- is equivalent to within 10 per cent to the
relation given in \citet{W02} for the now standard $\Lambda$CDM.

\paragraph*{Total gravitating mass and potential.}
Luminous and dark matter components combine to the total mass
density
\begin{equation}
\rho = \Upsilon \, \nu + \rho_\mathrm{DM},
\end{equation}
with $\rho_\mathrm{DM}$ being either $\rho_\mathrm{LOG}$ or $\rho_\mathrm{NFW}$.
The potential $\Phi$ follows by integrating Poisson's equation.

\subsection{Orbit collection}
\label{orbits}
Given $\Phi$, a large set of orbits is calculated in order to sample
the phase-space of the potential. Energies $E$ and angular momenta $L_z$ of the orbits
are chosen to connect every pair of equatorial radial grid bins by at least one equatorial
orbit. The surfaces of section (SOS) connected
to pairs of ($E,L_z$) -- here the upward orbital crossings with the equator --
are densely filled with orbits of all available shapes \citep{comadyn1}.
A typical orbit library contains between 6500 and
9500 orbits for $L_z > 0$. Each orbit's retrograde counterpart
with $L_z <0$ is included in the library by reversing the azimuthal velocities appropriately. 
In total a typical library then contains between 13000 to 19000 orbits.

\subsection{Orbit superposition}
\label{fit}
Any superposition of a library's orbits generates a model with a specific
internal density distribution and specific projected kinematics that can be compared with
the observations. The relative contribution of each 
orbit to the superposition -- the
orbital weight $w_i$ -- represents the total amount of light carried by 
orbit $i$. To fit a library to a given dataset the 
maximum entropy technique of \citet{maxspaper} is applied by maximizing
\begin{equation}
\label{maxs}
\hat{S} \equiv S - \alpha \, \chi^2_\mathrm{LOSVD},
\end{equation}
where 
\begin{equation}
\label{chilosvd}
\chi^2_\mathrm{LOSVD} \equiv \sum_{j=1}^{N_{\cal L}} \, \sum_{k=1}^{N_\mathrm{vel}} 
\left(
\frac{{\cal L}^{jk}_\mathrm{mod}-{\cal L}^{jk}_\mathrm{dat}}
{\Delta {\cal L}^{jk}_\mathrm{dat}}
\right)^2
\end{equation}
quantifies the deviations between the model LOSVDs
${\cal L}_\mathrm{mod}$ and the observed LOSVDs ${\cal L}_\mathrm{dat}$ at 
$N_{\cal L}$ spatial positions $j$ and in the $N_\mathrm{vel}$ bins of projected
velocity $k$. By $S$, the Boltzmann entropy
\begin{equation}
\label{bentropy}
S \equiv \int f \ln \left( f \right) \,
\der^3r \, \der^3v
\, = \sum_i w_i \ln \left( \frac{w_i}{V_i} \right)
\end{equation}
of the library's DF $f$ is denoted, $V_i$ is the orbital phase-volume of orbit $i$,
computed as in \citet{comadyn1}, and $\alpha$ is a regularization parameter 
(see Sec.~\ref{alpha}). 

The decomposition of the library potential $\Phi$ into two
components generated by the stellar and the
dark matter mass distributions, respectively, 
is meaningful only if the final orbit model self-consistently
generates the stellar contribution to the potential. Therefore,
the luminosity density $\nu$ is used as a boundary condition for the maximization of
equation (\ref{maxs}). This also guarantees a perfect match of the 
orbit model to the photometric observations.

\subsection{Comparing model with data kinematics}
\label{chivschi}
Since our original data set for NGC 4807 consists of Gauss-Hermite
parameters up to $H_4$ we will quote the deviations between model and data 
kinematics in terms of
\begin{eqnarray}
\label{chigheq}
\chi^2_\mathrm{GH} \equiv \sum_{j=1}^{N_{\cal L}} 
\left[
\left( 
\frac{v^j_\mathrm{mod} - v^j_\mathrm{dat}}{\Delta v^j_\mathrm{dat}} 
\right)^2 +
\left( 
\frac{\sigma^j_\mathrm{mod} - \sigma^j_\mathrm{dat}}{\Delta \sigma^j_\mathrm{dat}} 
\right)^2 + \right. \nonumber \\
\left.
\left( 
\frac{H^j_{3,\mathrm{mod}} - H^j_{3,\mathrm{dat}}}{\Delta H^j_{3,\mathrm{dat}}} 
\right)^2 +
\left( 
\frac{H^j_{4,\mathrm{mod}} - H^j_{4,\mathrm{dat}}}{\Delta H^j_{4,\mathrm{dat}}} 
\right)^2
\right].
\end{eqnarray}
The sum in equation (\ref{chigheq}) over $N_{\cal L}$ LOSVDs
contains $N_\mathrm{data} \equiv 4 \times N_{\cal L}$ terms and the parameters
$v_\mathrm{mod}$, $\sigma_\mathrm{mod}$, $H_{3,\mathrm{mod}}$ and $H_{4,\mathrm{mod}}$ are
obtained from the corresponding LOSVDs of equation (\ref{chilosvd})
by fitting a 4th-order Gauss-Hermite series\footnote{The fifth parameter of the 
fit, the intensity $\gamma_0$, is not included in equation (\ref{chigheq}) since we scale 
data as well as model LOSVDs to the same surface brightness before comparison.
This does not automatically imply that the fitted intensities $\gamma_0$ of model and 
data LOSVDs are identical, but it largely affects their differences. Following the
approach of \citet{G98} we therefore omit $\gamma_0$ in $\chi^2_\mathrm{GH}$.}.
Equation (\ref{chigheq}) can only be applied if the sampling of the LOSVD
is sufficient to get reliable and unbiased estimates of the Gauss-Hermite
parameters and it should be noticed that in the implementation of the orbit superposition
method applied here (in contrast to the 
programs building up on the work of \citealt{R97} and \citealt{Cre99})
$\chi^2_\mathrm{GH}$ is not explicitly minimized. We discuss the effects of using
equation (\ref{chigheq}) instead of equation (\ref{chilosvd}) to derive confidence regions 
in Sec.~\ref{chidisc}.

\section[]{Regularization}
\label{alpha}
The regularization parameter $\alpha$ in equation (\ref{maxs}) controls the relative
importance of entropy maximization and $\chi^2$-minimization in the model. Basically,
increasing $\alpha$ puts more weight on the $\chi^2$-minimization in the orbit superposition
and reduces the influence of $S$ in equation (\ref{maxs}). For data sets with relatively 
sparse spatial sampling as considered here, the freedom in the orbit 
superpositions allows to fit models to the noise in the data when applying $\alpha > 1$. Such 
orbit models show, however, large density depressions and contradict the traditional view of 
relaxed dynamical systems. On the other hand, models with lower $\alpha$ 
have smoother distribution functions in the sense that the adopted form for $S$ tends
to isotropize the orbital DF, thereby reducing its dependency on $L_z$ and $I_3$.
Fitting orbit models to data sets with different spatial coverage and quality will likely
change the effect of $\alpha$ on the final fit. Likewise, changing the functional form
of $S$ can be used to bias the models towards other than isotropic DFs \citep{comadyn1}.
The best choice for $\alpha$ and $S$ has to be investigated case-by-case, depending
on the galaxies under study and on the amount and form of information that is to be
extracted from the observations.

\subsection{Motivation}
\label{virtue}
\begin{figure}
\includegraphics[width=84mm,angle=0]{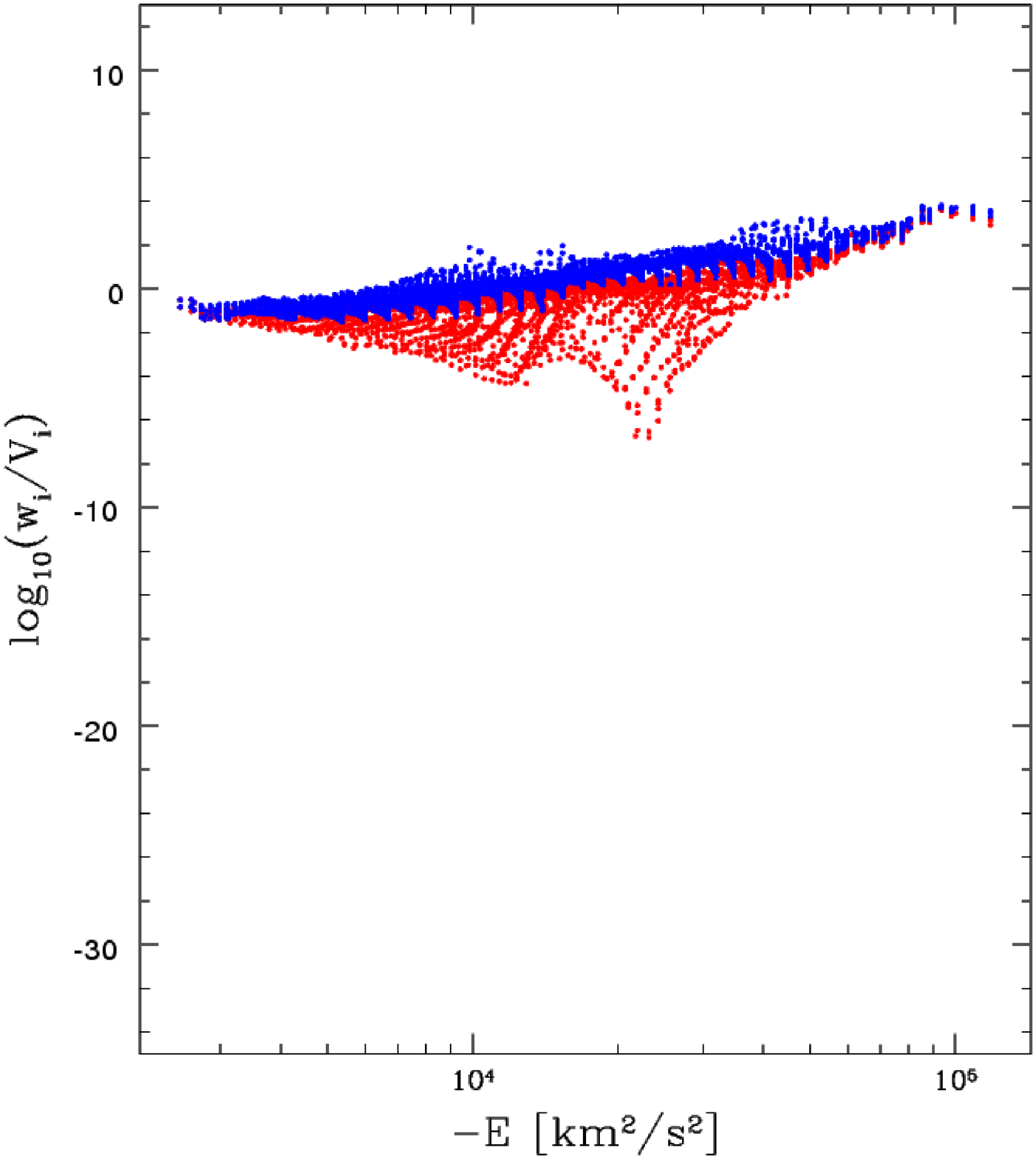}
\includegraphics[width=84mm,angle=0]{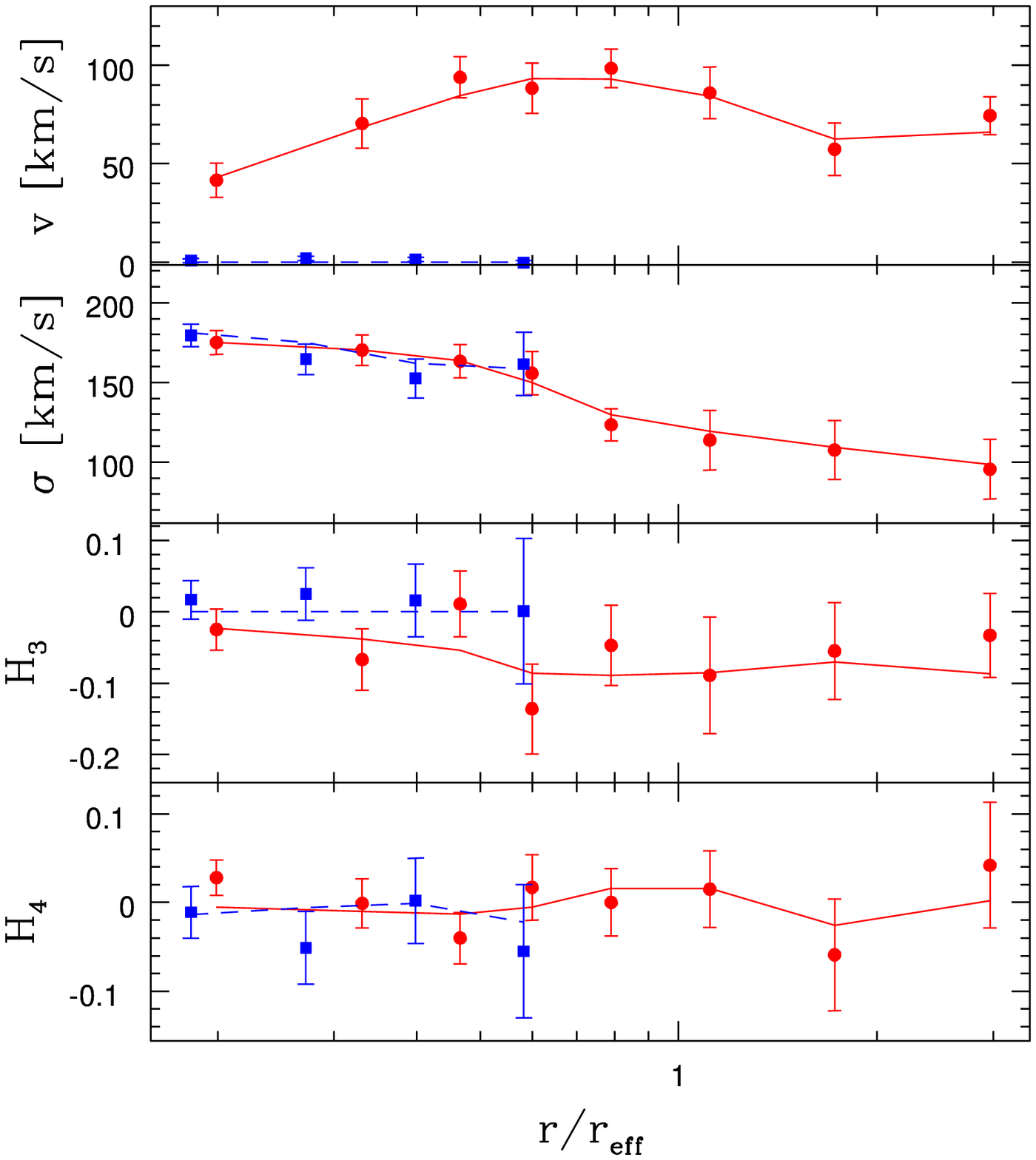}
\caption{Reconstructed phase-space densities of individual orbits against energy (upper
panel) for an isotropic rotator model of NGC 4807 (details in the text). For the underlying 
fit (lower panel; dots/squares: major/minor axis isotropic rotator model; solid/dashed 
lines: major/minor axis orbit model) a regularization of $\alpha = 0.02$ is used.}
\label{overfitting1}
\end{figure}

\begin{figure}
\includegraphics[width=84mm,angle=0]{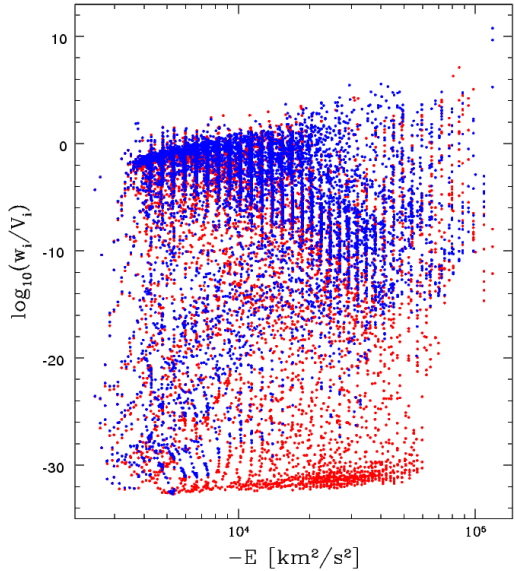}
\includegraphics[width=84mm,angle=0]{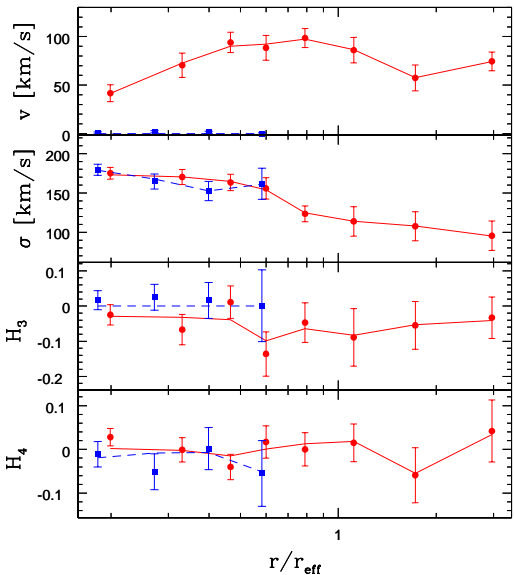}
\caption{As Fig.~\ref{overfitting1}, but for regularization with $\alpha = 6.73$.
Note that the lower boundary of phase-space densities in the upper panel is partly due to the 
program setting all orbital weights smaller than $w_i < w_\mathrm{min} \equiv 10^{-37}$
equal to $w_i \equiv w_\mathrm{min}$.}
\label{overfitting2}
\end{figure}

Figs.~\ref{overfitting1} and \ref{overfitting2} exemplify the effects of regularization
in terms of two distribution functions
reconstructed from fits of the same library for an edge-on isotropic rotator model 
of NGC 4807. The Gauss-Hermite profiles in the lower panels of the
figures are derived from velocity moments
obeying higher-order Jeans equations in the self-consistent potential of the deprojection
\citep{mag94}. Before calculating the Gauss-Hermite parameters the velocity moments are
slit averaged and seeing convolved to simulate the observations
of Sec.~\ref{kinematics}. Noise is added to $v$ and $\sigma$ according to the 
fractional errors of the observations and to $H_3$ and $H_4$ according to the 
absolute errors. The lower panels of Figs.~\ref{overfitting1} and \ref{overfitting2},
respectively, display the rotator kinematics along the major and minor axes,
together with fits of an orbit library containing $2 \times 6922$ orbits. For 
Fig.~\ref{overfitting1} the fits were obtained at $\alpha=0.02$ and for 
Fig.~\ref{overfitting2} at $\alpha=6.73$. The distribution functions reconstructed from the 
two fits are plotted against orbital energy $E$ in the upper panels of the figures. 
Each dot represents the phase-space density $w_i/V_i$ along a 
single orbit, scaled according to $\sum w_i = \sum V_i = 1$ \citep{comadyn1}.
The more regularized fit in Fig.~\ref{overfitting1} yields 
$\chi^2_\mathrm{GH}/N_\mathrm{data} = 0.3$ ($N_\mathrm{data}=48$), while for 
Fig.~\ref{overfitting2} the goodness of fit is 
$\chi^2_\mathrm{GH}/N_\mathrm{data} = 0.17$. 

\begin{figure}
\includegraphics[width=84mm,angle=0]{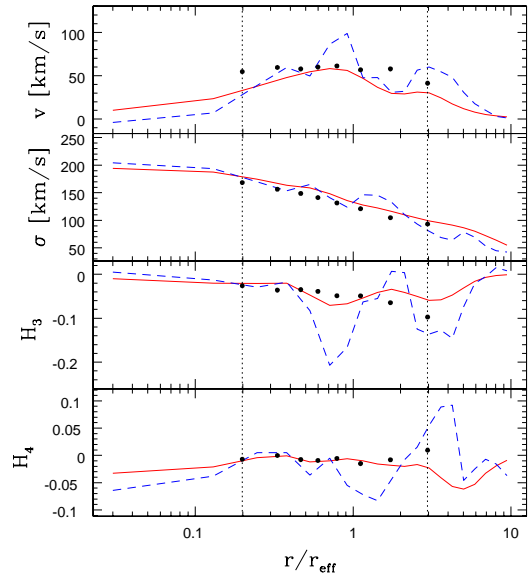}
\caption{Projected kinematics of the two DFs of Figs.~\ref{overfitting1} and 
\ref{overfitting2} along a diagonal axis with position angle $\vartheta=45\degr$. 
Solid lines: $\alpha=0.02$ ($\rightarrow$ Fig.~\ref{overfitting1}); dashed lines: 
$\alpha=6.73$ ($\rightarrow$ Fig.~\ref{overfitting2}); dots: kinematics of the isotropic 
rotator model. Note that the orbit superpositions were fitted to the model only along the 
major and minor axes.}
\label{projkin}
\end{figure}

The DF at $\alpha = 6.73$ has density depressions of several orders of 
magnitude for orbits with roughly the same energy. 
Orbit models with such DFs are difficult to interpret as the result of 
relaxation processes like violent relaxation and others, occurring in the 
dynamical evolution of real galaxies (e.g. \citealt{Lyn67}). They often predict uncommon
kinematics along position angles not covered by observational data. To illustrate this,
the projected kinematics predicted along a diagonal axis (position
angle $\vartheta = 45 \degr$) by the two DFs of Figs.~\ref{overfitting1} and 
\ref{overfitting2} are plotted in Fig.~\ref{projkin}. For comparison the (undisturbed) 
kinematics of the isotropic rotator model are also overplotted. The
profiles of the almost non-regularized ($\alpha=6.73$) fit show large point-to-point
variations, which cannot easily be reconciled with the scatter in real observations.
Also, the mean deviation between
the unregularized orbit superposition and the input isotropic rotator (IR) model are larger 
for the non-regularized one than for the regularized one, e.g. 
$\langle H_3^\mathrm{IR}-H_3\rangle = 0.06$ for $\alpha=6.73$ compared with 
$\langle H_3^\mathrm{IR}-H_3\rangle = 0.02$ for $\alpha=0.02$.

The above example demonstrates the importance of regularization in the 
context of recovering internal dynamics of isotropic rotators 
from sparse data sets typical for our
sample of Coma galaxies. For NGC 4807 we have determined the optimal amount of regularization
by means of Monte-Carlo simulations of several such 
isotropic rotator models (see Sec.~\ref{irot}
below). 

The DFs of isotropic rotator models are of the form $f=f(E,L_z)$ and represent
only a minority of all possible DFs, since they are constant along $I_3$.
The choice of such reference models for galaxies like 
NGC 4807 (fast rotating, faint giant ellipticals) is observationally motivated 
\citep{Kor96} and further supported by
the weak velocity anisotropy found in previous dynamical studies
(e.g. \citealt{G01}; see also Sec.~\ref{kin5975} for the case of NGC 4807), implying
only a mild dependence of the DF on $I_3$. For more
luminous ellipticals (shaped primarily by velocity anisotropy) or lenticulars embedded in
roundish halos other reference models should be explored. We will turn to this in a 
future publication.

\subsection{Regularization from isotropic rotator models}
\label{irot}
The isotropic rotator models constructed to determine the optimal regularization with respect 
to the analysis of NGC 4807's internal structure are based
on the edge-on as well as the $i=50\degr$ deprojections.
For numerical reasons, however, the models at $i=50\degr$ are forced to have
isophotes close to pure ellipses by setting $a_4 \equiv a_6$ for the deprojection.
Six models have been probed, three
at each of the inclinations $i=50,90\degr$: one self-consistent, one embedded in a LOG-halo 
and a third embedded in a NFW-halo. Kinematic profiles are calculated from higher order 
internal velocity moments \citep{mag94} following the procedure described in 
Sec.~\ref{virtue}. Orbit libraries are fitted to the kinematic profiles and the internal 
velocity moments reconstructed from the fits are compared with the original input-moments 
for various values of $\alpha$. The optimal balance between
entropy maximization and $\chi^2$-minimization is revealed where the reconstructed 
internal velocity moments are closest to the input model.

\begin{figure}
\includegraphics[width=84mm,angle=0]{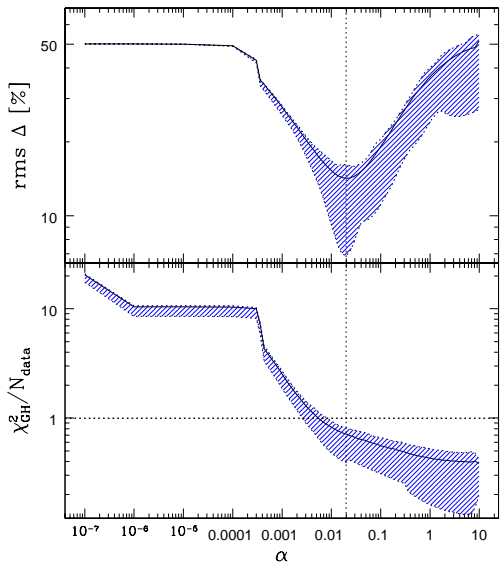}
\caption{Match $\Delta(\alpha)$ of internal velocity moments and 
$\chi^2_\mathrm{GH}/N_\mathrm{data}$ of the Gauss-Hermite parameters as a function
of the regularization parameter $\alpha$. Lines represent the mean over 60 fits to Monte-Carlo
realizations of the isotropic rotator's kinematic profiles; shaded regions encompass 68 per
cent of the simulations closest to the corresponding minimum in $\Delta(\alpha)$ and
$\chi^2_\mathrm{GH}/N_\mathrm{data}$, respectively.}
\label{rmssci90}
\end{figure}

Fig.~\ref{rmssci90} shows the results for a self-consistent model based on the 
$i=90\degr$-deprojection. The upper panel shows the rms difference $\Delta(\alpha)$ 
between original and 
reconstructed internal velocity moments (up to second order) as a function of the
regularization parameter $\alpha$. The rms is evaluated between the innermost and 
outermost fitted data points, at all position angles $\vartheta$, including intermediate 
ones, not covered by data points. The lower panel illustrates the goodness of fit 
$\chi^2_\mathrm{GH}/N_\mathrm{data}$. Solid lines correspond to the mean obtained by 
fitting orbit models to 60 Monte-Carlo realizations of the simulated data. 
Shaded areas comprise the 68 per cent fraction of
best reconstructions (upper panel) and 68 per cent fraction of best fitting solutions 
(lower panel), respectively.

The best results are obtained for $\alpha = 0.02$, when internal kinematics of the fits
follow the input moments to an accuracy of about 15 per cent in the mean. 
Lower $\alpha$ yield less accurate reconstructions
since the orbit superpositions do not fit the data enough (see $\chi^2_\mathrm{GH}$ in 
the lower panel). For larger $\alpha$ on the other hand, the rms $\Delta(\alpha)$ 
increases, because the library starts to fit the noise in the data.

On average, the orbit models at $\alpha=0.02$ fit with
$\chi^2_\mathrm{GH}/N_\mathrm{data} < 1$. A proper normalization for $\chi^2_\mathrm{GH}$ 
is however hard to obtain, since we do not exactly know the number of free parameters in the 
library. For $\alpha=0$ this number is zero, because $\chi^2$ does not appear in 
equation (\ref{maxs}) while for $\alpha \rightarrow \infty$ there are roughly as many 
free parameters as orbits in the library\footnote{We have neglected the boundary 
condition related to $\nu$. This is expected to reduce the number of free parameters,
but is unlikely to break its dependency on $\alpha$.}, because then $S$ becomes negligible 
in equation (\ref{maxs}). Thus, we do not try to calculate a reduced $\chi^2$.

Table~\ref{alstar} summarizes the total of simulations done to fix $\alpha$. The first 
column labels the three models investigated at each of the two inclinations $i=50\degr$ and 
$i=90\degr$ (SC: self-consistent models without halo; LOG: a dark halo with 
$r_c = 5 \, \mathrm{kpc}$ and $v_c = 265 \, \mathrm{km/s}$ according to equation 
(\ref{nis}) is added to the stars; NFW: dark halo profile like in equation (\ref{nfw}) 
with $r_s = 10.0 \, \mathrm{kpc}$, $c=27.0$; for all models $\Upsilon=3.0$). 
The halo models are
constructed to produce a roughly flat circular velocity. In the second column
of Table~\ref{alstar} the regularization $\alpha_0$ is quoted attaining the best
reconstruction of the internal velocity moments, quantified in the third column as
$\Delta(\alpha_0) = \min (\Delta(\alpha))$. The corresponding goodness of fit
$\chi^2_\mathrm{GH}(\alpha_0)$ is given in the fourth column of the table. 
The number of orbits in each of the fitted libraries is found in the last column.

\begin{table}
\begin{tabular}{@{}lcccc@{}}
\hline
input model & $\alpha_0$ & $\Delta(\alpha_0)$ & $\chi^2_\mathrm{GH}(\alpha_0)$ & $N_\mathrm{orbit}$ \\
\hline
$i=90\degr$, SC & $0.0199$ & $13.5 \, \%$ & $0.712$ & $2 \times 6922$ \\
$i=90\degr$, LOG & $0.0166$ & $12.7 \, \%$ & $0.719$ & $2 \times 8273$ \\
$i=90\degr$, NFW & $0.0138$ & $13.2 \, \%$ & $0.692$ & $2 \times 8469$ \\
$i=50\degr$, SC & $0.0238$ & $12.6 \, \%$ & $0.539$ & $2 \times 6697$ \\
$i=50\degr$, LOG & $0.0166$ & $13.8 \, \%$ & $0.826$ & $2 \times 8126$ \\
$i=50\degr$, NFW & $0.0166$ & $13.5 \, \%$ & $0.797$ & $2 \times 8324$ \\
\hline
\end{tabular}
\caption{Summary of simulations aimed at optimizing the regularization parameter $\alpha$.
Columns from left to right: inclination of reference model; potential of reference model
(SC = self-consistent, LOG = logarithmic spherical halo, NFW = spherical
NFW-halo, details in the text); regularization parameter $\alpha_0$ that yields the 
best reconstruction of internal velocity moments; rms $\Delta(\alpha_0)$ between internal 
velocity moments of reference models and reconstructions achieved with $\alpha=\alpha_0$; 
goodness of fit $\chi^2_\mathrm{GH}(\alpha_0)$; number of 
orbits $N_\mathrm{orbit}$ used for the modeling.}
\label{alstar}
\end{table}

On average, the six probed models yield $\langle \alpha_0 \rangle = 0.0179 \pm 0.0035$.
The low scatter seems to indicate that $\alpha_0$ does not depend strongly on the potentials
tested. In the implementation of the maximum entropy technique used here, $\alpha$ is 
increased iteratively in discrete steps (see Richstone et al., in preparation). 
For the models of NGC 4807 we apply $\alpha=0.0199 \approx 0.02$, which is the closest
larger neighbor of $\langle \alpha_0 \rangle$ in these iterations.

In the remainder of the paper, by quoting 68 (90, 95) per cent confidence levels, we
always refer to all orbit models whose $\chi^2_\mathrm{GH}/N_\mathrm{data}$ are below
the corresponding maximum $\chi^2$-level of the 68 (90, 95) per cent best-matching
fits in the simulations. For these fits parameters like $\Upsilon$, $i$,
$r_c$, $v_c$ \ldots are set equal to the true input values, but in the analysis of 
real galaxies they are varied. Strictly speaking then, the $\chi^2$-statistics from 
the simulations are not directly applicable to real galaxies. 
For example, to determine the correct 
statistics for the case where $\Upsilon$, $r_c$ and $v_c$ are varied 
to find the best-fitting mass-model, about $10^2$ orbital 
fits for each triple of ($\Upsilon,r_c,v_c$) are necessary to yield the corresponding 
three dimensional $\chi^2$-distribution. Such simulations however cannot be performed in 
a reasonable amount of time with present-day computer power.

Probably, our
confidence regions derived as described above overestimate the error budget. 
For example, in the simulations we find 68 per cent of all orbits within 
$\Delta \chi^2_\mathrm{GH}/N_\mathrm{data} = 0.38$ from the mean best-fit values. 
In contrast, applying classical $\Delta \chi^2$-statistics for a two parameter fit, 
yields about 95 per cent confidence at the same 
$\Delta \chi^2_\mathrm{GH}/N_\mathrm{data} = 0.38$.

\section[]{Recovering isotropic rotator models}
\label{confidence}
The aim of this Section is to quantify in a practical sense to which degree a sparse data set
like the one described in Sec.~\ref{kinematics} constrains the internal kinematics and
mass structure of a typical galaxy in our sample.

\subsection{Mass-to-light ratio $\Upsilon$ and inclination $i$}
\label{mlrecon}
\begin{figure}
\includegraphics[width=84mm,angle=0]{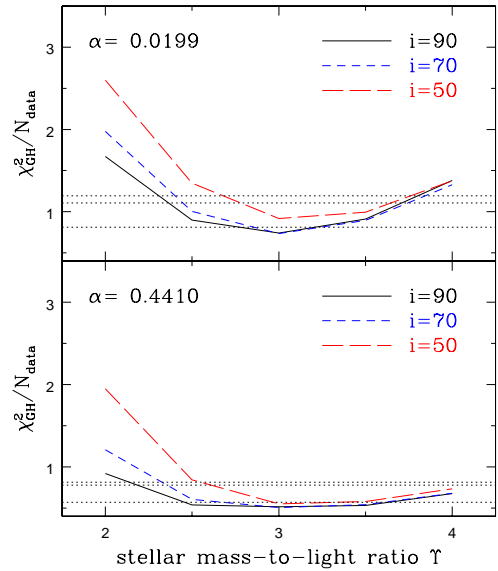}
\caption{$\chi^2_\mathrm{GH}$ per data point ($N_\mathrm{data}=48$) as function of stellar 
$M/L$ for three different inclinations $i$. Top panel: with optimal smoothing, bottom panel: 
weak smoothing. The input 2I-model has $M/L=3.0$ and is edge-on. Pointed horizontal lines
in each panel represent the $\chi^2_\mathrm{GH}$ values enclosing 68, 90 and 95 per cent
of the Monte-Carlo simulations, respectively.}
\label{chiml.re}
\end{figure}

First, we disregard the possible presence of a dark halo and try to
recover the mass-to-light ratio and inclination of the self-consistent edge-on model
in Table~\ref{alstar}. Therefore, Fig.~\ref{chiml.re} resumes the results of fitting 
libraries with mass-to-light ratios in the range $2.0 \leqslant \Upsilon \leqslant 4.0$ and 
inclinations $i=50,70,90\degr$ to the kinematics of this isotropic rotator model.
The goodness of fit $\chi^2_\mathrm{GH}/N_\mathrm{data}$ for each pair
($\Upsilon,i$) is averaged over 10 Monte-Carlo realizations of the kinematic profiles and
evaluated at two different regularizations. 
For the upper panel $\alpha=0.02$ according 
to the simulations described in Sec.~\ref{irot}. Horizontal dotted lines in the plot 
correspond to 68, 90 and 95 per cent confidence levels as derived from the statistics of the 
simulations in Sec.~\ref{irot}. 

One can read from the plot that the mass-to-light ratio is well recovered: the orbital
fits reveal $\Upsilon = 3.0 \pm 0.5$, where the input model has 
$\Upsilon_\mathrm{IR} = 3.0$. The minimum $\chi^2_\mathrm{GH}$ occurs at the true
value independent of the assumed inclination. The latter is only weakly constrained. 
For $\Upsilon = 3.0$ an inclination of $i=70\degr$ results in nearly the same 
$\chi^2_\mathrm{GH}/N_\mathrm{data}$ as the fit with the input $i_\mathrm{IR}=90\degr$. 
Even the orbit models with $i=50\degr$ can only be rejected with less than 
90 per cent confidence.

For the lower panel the fits are evaluated at $\alpha=0.44$ (this choice of $\alpha$ is
motivated in Sec.~\ref{mass5975}). Although at this less
restrictive regularization the confidence intervals shrink as compared with $\alpha=0.02$ the
mass-to-light ratio is now less constrained, $\Upsilon = 3.0 \pm 1.0$. The same holds for
the inclination: at $\Upsilon = 3.0$ all three probed inclinations are now within the
68 per cent confidence interval.

\subsection{Internal kinematics}
\begin{figure}
\includegraphics[width=84mm,angle=0]{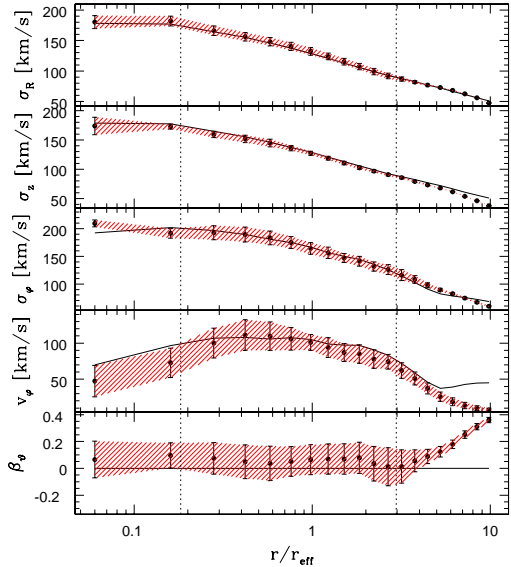}
\caption{Internal velocity moments along the major axis of libraries fitted to the 
self-consistent edge-on isotropic rotator model of Sec.~\ref{virtue}. Dots: 
mean from fits to 60 Monte-Carlo realizations of the kinematic data; shaded regions: 
$1 \, \sigma$ deviation from the mean; solid lines: velocity moments of the input model;
dotted lines: boundaries of the radial region included in the fit.}
\label{intmomsci90}
\end{figure}

To give an impression about the ability to recover internal motions (assuming
the potential is known) Fig.~\ref{intmomsci90} compares internal velocity moments 
reconstructed from libraries fitted to
the self-consistent edge-on isotropic rotator model of Sec.~\ref{virtue}
with the corresponding moments of the input model. Solid lines portray the isotropic rotator
model and points show the average reconstructed moments from fits to 60 Monte-Carlo 
simulations of the kinematic profiles; error bars indicate $1 \, \sigma$ deviations from 
the mean. For the reconstructions $\alpha = 0.02$ is used.

The three upper rows demonstrate that for the internal major axis the second order moments 
$\sigma_R$, $\sigma_z$ and $\sigma_\varphi$ are accurately reproduced by the fit 
over the region between the pointed vertical lines indicating the radii of 
the innermost and outermost kinematical points included in the fit. The fractional errors 
of $\sigma_R$ and $\sigma_\varphi$ are below 3 per cent and of $\sigma_z$ are below 6 per 
cent. For the internal rotation velocity $v_\varphi$ the fractional errors are larger 
than 10 per cent at some points. The match of the rotation velocity can be increased when
going to larger $\alpha$, but then the second order moments start to wiggle around the
input moments and the overall match of reconstructed to original moments becomes
worse. Fig.~\ref{intmomsci90} is representative also for the remaining internal 
position angles of the libraries. It follows that most of the rms in Fig.~\ref{rmssci90} 
results from a mismatch in $v_\varphi$.

The last row shows a comparison of the anisotropy parameter
$\beta_\vartheta = 1 - \sigma_\vartheta^2/\sigma_r^2$, which vanishes for isotropic rotator 
distribution functions $f=f(E,L_z)$. The reconstructions are consistent with 
$\beta_\vartheta = 0$ given the scatter caused by the noisy data. The small
offset of $\Delta \beta_\vartheta = 0.05$ is due to a slight overestimation of the radial
velocity dispersion and emphasizes how sensitive $\beta_\vartheta$ 
responds to small inaccuracies in
the velocity dispersions.

\subsection{Mass distribution}
\label{halop}
\begin{figure}
\includegraphics[width=84mm,angle=0]{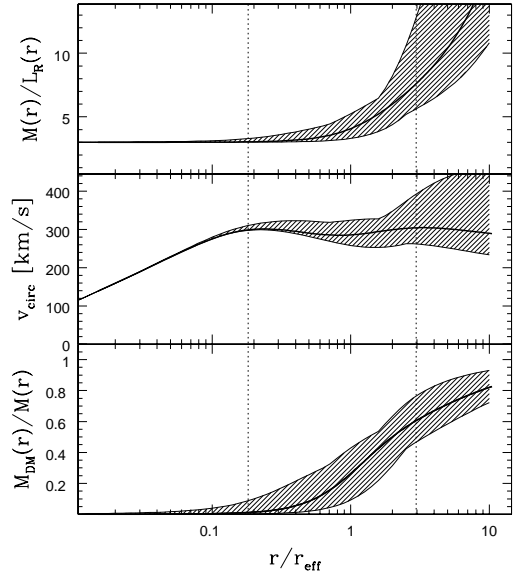}
\caption{Accuracy of the mass reconstruction. Thick lines: the isotropic 
rotator model marked by the asterix in Fig.~\ref{nisconthalo}; 
shaded areas: 68 per cent confidence regions of 
fits to pseudo data sets of the model. For each mass-model the goodness-of-fit is
averaged over 10 fits to MC-realizations of pseudo-data.}
\label{halorecon}
\end{figure}

The next step is to recover the structure of the isotropic rotator model in the
second row of Table~\ref{alstar}, where a logarithmic dark halo is present. 
To this end, we fitted libraries with different
dark halos to pseudo data sets of the rotator model, keeping the stellar mass-to-light 
ratio constant. Fig.~\ref{halorecon} combines the results of the simulations.
It shows, from top to bottom, cumulative mass-to-light ratio $M(r)/L(r)$, circular velocity
$v_\mathrm{circ}(r)$ and dark matter fraction $M_\mathrm{DM}(r)/M(r)$ as a function
of radius. Vertical dotted lines mark the boundaries of the spatial region covered with
kinematic data, thick lines display the input model. The shaded areas have been constructed by
determining at each radius the minimum and maximum of $M/L$, $v_\mathrm{circ}$ and 
$M_\mathrm{DM}/M$ of all libraries within the 68 per cent confidence range derived from the 
simulations of Sec.~\ref{irot}. Libraries with LOG-halos as well as NFW-halos have been
tried (see below) and for each library fits to 10 realizations of the isotropic rotator
kinematics were averaged. 

The figure demonstrates that in the region covered by the data the mass structure of 
the input model is well reproduced, with progressively
larger scatter towards the outer edge of the data. The uncertainty of the mass-to-light
ratio is $\Delta M/L \approx 1.0$ at $1 \, r_\mathrm{eff}$, the circular velocity
is accurate to 10 per cent at the same central distance. In the outer parts orbit
superpositions with larger masses than the input model are better consistent with the data 
than smaller mass models. Nevertheless, the dark matter fraction is determined to 
$\Delta (M_\mathrm{DM}/M) = 0.2$ even at the outermost data point. 

\begin{figure}
\includegraphics[width=84mm,angle=0]{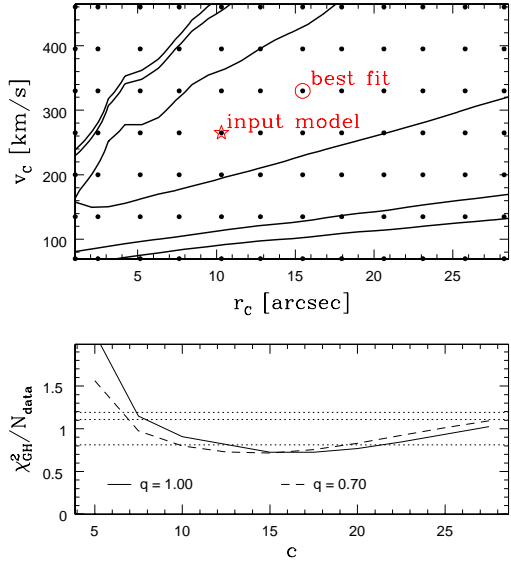}
\caption{Recovering an isotropic rotator model's dark halo: $r_c$ and $v_c$ of the input model
model are marked by the asterix in the top panel, the best-fit reconstruction 
is marked by the circle. Lines in the top panel display 68, 90 and 95 per cent confidence 
contours of the fits. The lower panel shows -- for the same input model -- $\chi^2$ of 
NFW-type fits (solid/dashed: $q=1.0$/$q=0.7$); 
horizontal lines: 68, 90 and 95 per cent confidence levels. Each library
has been fitted to 10 realizations of the input-model kinematics.}
\label{nisconthalo}
\end{figure}

The top panel of Fig.~\ref{nisconthalo} displays 68, 90 and 95 per cent confidence intervals 
for the two parameters $r_c$ and $v_c$ of LOG-halos (cf. equation \ref{nis}). 
Each dot marks a pair of ($r_c,v_c$) probed by fitting a library to 10 realizations of the
pseudo-data as described above.
The input model's dark halo parameters are marked by the asterix, the circle designates
the best-fitting orbit model. 

As expected, the halo parameters are not well constrained. Although the best-fitting 
parameter pair is close to the input model, the 68 per cent confidence contour comprises
a large set of libraries and remains open to the upper right edge of the plot. This follows
from a degeneracy between the two parameters $r_c$ and $v_c$ (e.g. \citealt{G98}). Increasing
$r_c$ and $v_c$ appropriately puts more mass in the outer parts of the halos while rendering
the regions covered by kinematics roughly unchanged. The increasing width of the shaded 
areas in the upper two panels of Fig.~\ref{halorecon} is an illustration of this degeneracy.

We also fitted one parameter NFW-halos according to equation (\ref{family}) to the same 
isotropic rotator model in order to examine whether these profiles can be excluded by the
data set at hand. The lower panel of Fig.~\ref{nisconthalo} displays the results in terms
of $\chi^2_\mathrm{GH}/N_\mathrm{data}$ as a function of concentration index $c$. Horizontal 
dotted lines correspond to 68, 90 and 95 per cent confidence levels, the solid line
shows fits with spherical halos ($q=1.0$), while for the dashed line $q=0.7$.
Apparently, neither the halo flattening nor the central halo density slope are constrained 
by the data, because both spherical as well as flattened NFW-halos exist 
that provide fits equally as good as the LOG-halos. These NFW-halos join smoothly to
the mass-distribution recovered in Fig.~\ref{halorecon} and mimic LOG-halos over the
limited spatial region covered by kinematic data.

Fig.~\ref{nisconthalo} suggests that it is not possible to discriminate between
LOG and NFW halos insofar as they provide similar mass distributions over the spatial region 
sampled by kinematic data. More extended data sets are likely to reduce this degeneracy, but 
will probably not completely remove it. On the other hand, as Fig.~\ref{halorecon}
reveals, the mass distribution and composition of the input model can be recovered well, 
independently of the specific parameterization chosen to emulate it.

\section[]{Dark matter in NGC 4807}
\label{mass5975}
Now, we turn to the analysis of the distribution of luminous and dark mass in NGC 4807.
Fig.~\ref{chimlbest} shows the minimum $\chi^2_\mathrm{GH}/N_\mathrm{data}$ 
(scaled, see below) at each modelled stellar mass-to-light ratio 
$\Upsilon \in \{1.0,2.0,2.5,3.0,3.5,4.0,5.0\}$. Horizontal lines indicate the 68, 90 and 95 
per cent confidence limits derived from the simulations of Sec.~\ref{irot}. For the upper 
panel $\alpha=0.02$ is used and LOG-potentials with stellar mass-to-light ratios 
$2.5 \leqslant \Upsilon \leqslant 3.0$ fit best to the data. Projected 
kinematics of the best-fitting orbit model are plotted in the upper panel of 
Fig.~\ref{ghfitbest} together with the data. The model is edge-on, 
has stellar $\Upsilon=3.0$ and a spherical dark halo of LOG-type with core radius 
$r_c=6.8 \, \mathrm{kpc}$ and asymptotic circular velocity $v_c = 300 \, \mathrm{km/s}$. 
It fits the data very well with (unscaled)
$\min (\chi^2_\mathrm{GH}/N_\mathrm{data})=0.17$, even too well when compared with the
average $\langle \chi^2_\mathrm{GH}/N_\mathrm{data}\rangle=0.71$ expected from the 
Monte-Carlo simulations. Since both, the scatter in the kinematic data from 
different sides of the galaxy (cf. Sec.~\ref{kinematics}) and the point-to-point variations 
in the profiles are smaller than the statistical error bars, the latter 
might be slightly overestimated. Hence, in the
following all $\chi^2_\mathrm{GH}/N_\mathrm{data}$ are rescaled such that
$\min (\chi^2_\mathrm{GH}/N_\mathrm{data}) = \langle \chi^2_\mathrm{GH}/
N_\mathrm{data}\rangle$. The good match of model and data nevertheless reconfirms that 
NGC 4807 is consistent with axisymmetry.

Applying regularization with $\alpha = 0.02$, self-consistent models (mass follows light) are 
ruled out by more than $95$ per cent confidence (see thin solid
line in the upper panel of Fig.~\ref{chimlbest}). The best-fitting
self-consistent model fits with (rescaled) $\chi^2_\mathrm{GH}/N_\mathrm{data}=1.67$ 
and is compared with the data in the bottom
panel of Fig.~\ref{ghfitbest}. It has $\Upsilon = 3.5$, larger than the best-fitting halo
model, but fails to reach the measured rotation in the outer parts of the galaxy.
A drop in the outer major axis $H_3$-profile indicates that the model maintains
the measured dispersion $\sigma$ around the observed $v$ partly by putting light on
retrograde orbits. This causes the LOSVD to fall off too sharply at velocities
larger than $v$ and to develop a wing at negative velocities, resulting in the too negative
$H_3$. Lack of support for dispersion at large rotation velocities 
hints at mass missing in the outer parts
of this model. Higher mass-to-light ratios, however, would strengthen the mismatch between
the central regions of the model and the innermost (minor axis) velocity dispersions.

\begin{figure}
\includegraphics[width=84mm,angle=0]{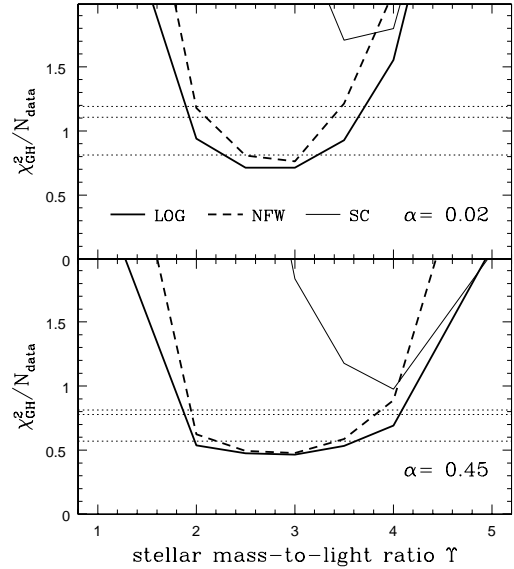}
\caption{Deviations between the observed kinematics of NGC 4807 and the
fitted libraries in terms of $\chi^2_\mathrm{GH}$ as a function of $\Upsilon$.
At each $\Upsilon$ the minimum $\min(\chi^2_\mathrm{GH})$ is plotted, separately for the
case of LOG-halos (thick solid line), one-parameter NFW-halos (thick dashed) and for the 
self-consistent case (thin solid). For the top row
the regularization parameter is $\alpha = 0.02$, for the bottom row
$\alpha = 0.45$.}
\label{chimlbest}
\end{figure}

To push the self-consistent model to the same level of agreement with the data reached
in the upper panel of Fig.~\ref{ghfitbest}, regularization must be lowered to 
$\alpha=0.45$. Although the best-fitting self-consistent
model then gives a satisfactory fit in the sense of a reasonable 
$\chi^2_\mathrm{GH}/N_\mathrm{data}$, it is again ruled out by more than $95$ per cent 
confidence when compared with the mean $\langle \chi^2_\mathrm{GH}/N_\mathrm{data} \rangle$ 
using the Monte-Carlo simulations. This, because
halo models at $\alpha=0.45$ still give significantly better fits to the data. The 
only effect of lower regularization -- as shown in the run of 
$\chi^2_\mathrm{GH}/N_\mathrm{data}$ versus $\Upsilon$ in the lower panel of 
Fig.~\ref{chimlbest} -- is a slight broadening of the minimum region in  
$\chi^2_\mathrm{GH}/N_\mathrm{data}$ and, consequently, an expansion of the allowed range
of mass-to-light ratios, yielding now $2.0 \leqslant \Upsilon \leqslant 3.5$ as compared with 
$2.5 \leqslant \Upsilon \leqslant 3.0$ at $\alpha=0.02$. Concluding, 
NGC 4807 cannot be convincingly modelled with 
self-consistent orbit models, even at weak regularization.

\begin{figure}
\includegraphics[width=84mm,angle=0]{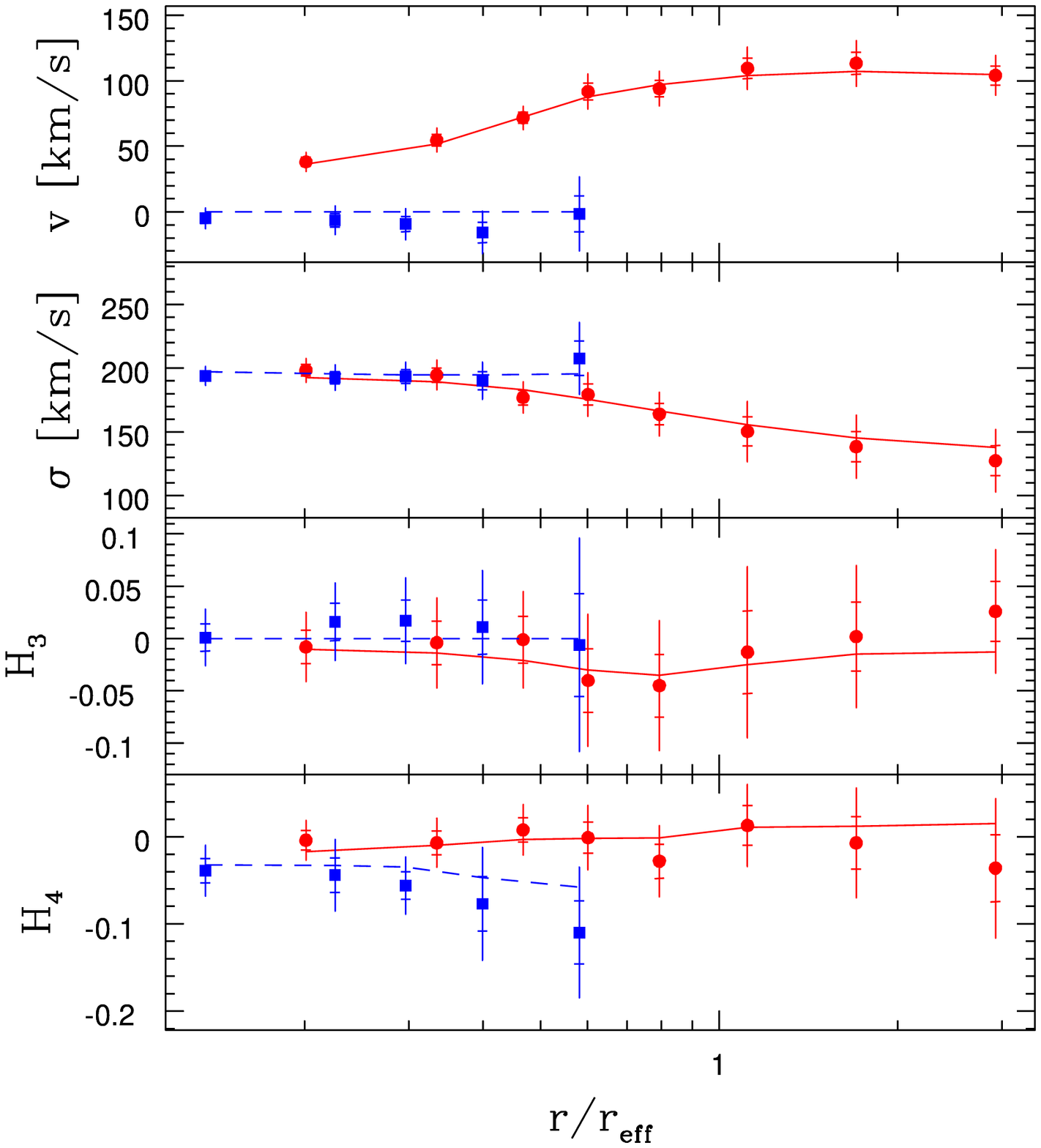}
\includegraphics[width=84mm,angle=0]{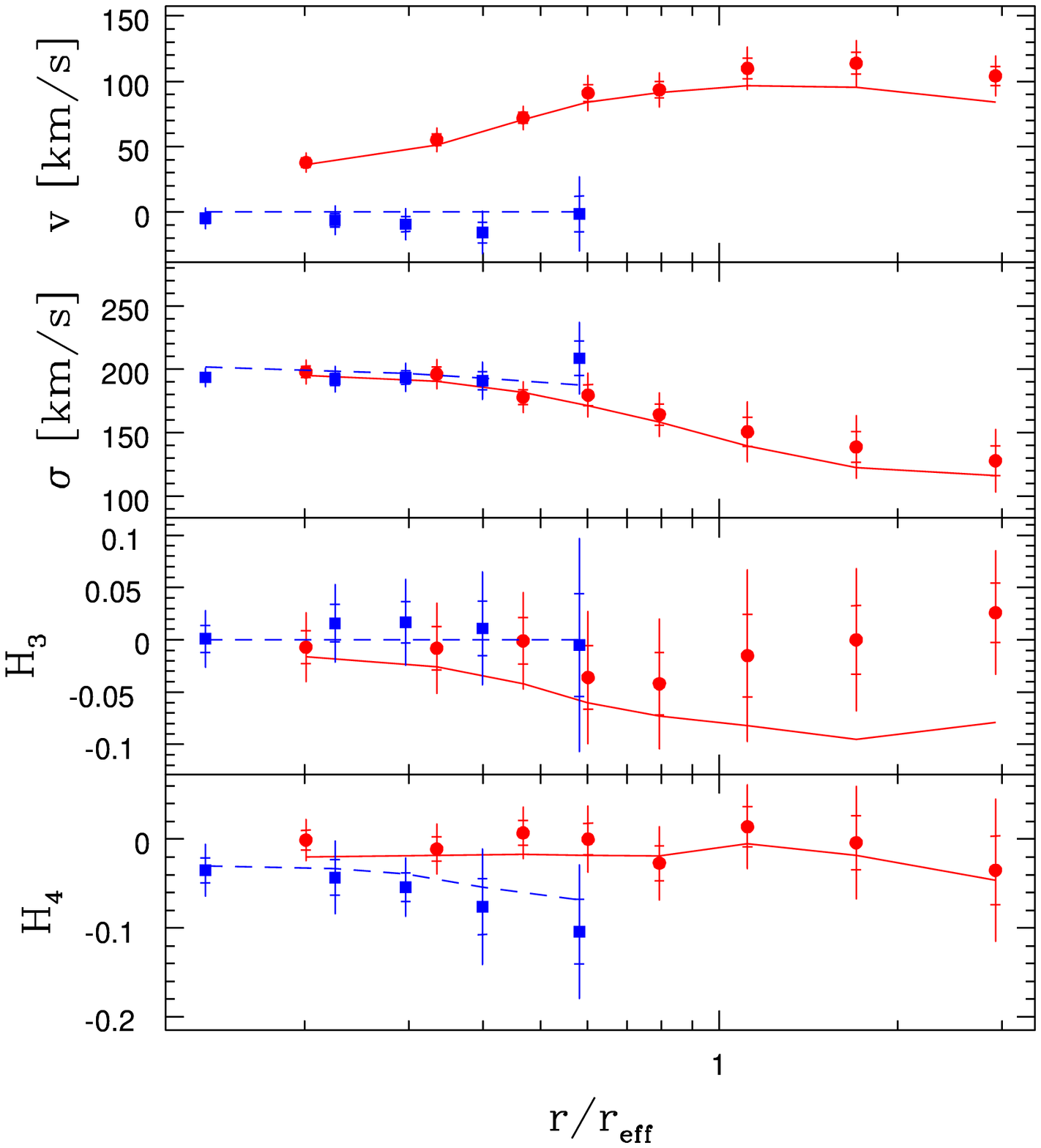}
\caption{Comparison of NGC 4807 (dots: major axis kinematic data; squares: minor axis) 
with the overall best-fitting model (upper panel) and best-fitting self-consistent
model (lower panel). Both orbit models are at $i=90\degr$. For each data point two error
bars are given: the statistical error from the observational data (larger, without marks) and
the error bar corresponding to the scaled $\chi^2_\mathrm{GH}$ (marks).}
\label{ghfitbest}
\end{figure}

\begin{figure}
\includegraphics[width=84mm,angle=0]{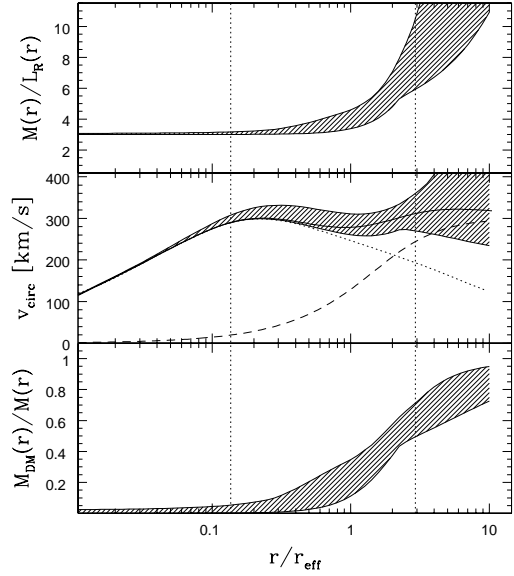}
\caption{Mass-structure of NGC 4807. Shaded regions are constructed as for 
Fig.~\ref{halorecon}. In the middle panel, the circular velocity curve of the 
best-fitting orbit model (solid line) and its decomposition into the stellar 
(dotted) and dark matter part (dashed) are displayed.}
\label{massstruc}
\end{figure}

In Fig.~\ref{massstruc} (from top to bottom) the 68 per cent confidence regions of
cumulative mass-to-light ratio $M(r)/L_R(r)$, circular velocity $v_\mathrm{circ}$
and dark matter 
fraction $M_\mathrm{DM}(r)/M(r)$ are shown (analogue to Fig.~\ref{halorecon}).
Within 1 $r_\mathrm{eff}$ dark matter is negligible and
the dynamical mass-to-light ratio equals the stellar one, $M(r)/L_R(r)=\Upsilon=3.0$. Between
$1 \, r_\mathrm{eff}$ and $3 \, r_\mathrm{eff}$ dark matter comes up from 
$M_\mathrm{DM}/M=0.21 \pm 0.14$ to $M_\mathrm{DM}/M=0.63 \pm 0.13$
and combines with the luminous matter
to a roughly flat circular velocity curve with $v_\mathrm{circ} = 280 \pm 30$ km/s at
$1 \, r_\mathrm{eff}$ and $v_\mathrm{circ} = 318 \pm 48$ km/s at 
the last kinematic data point.
In the same spatial region the total mass-to-light ratio rises from
$M(r)/L_R(r)=3.8 \pm 0.8$ to $M(r)/L_R(r)=8.4 \pm 2.4$.
Dark mass equals luminous mass at roughly two effective radii.
Beyond the last kinematic data point, the models are not well constrained and the profiles
start to diverge. 

\begin{figure}
\includegraphics[width=84mm,angle=0]{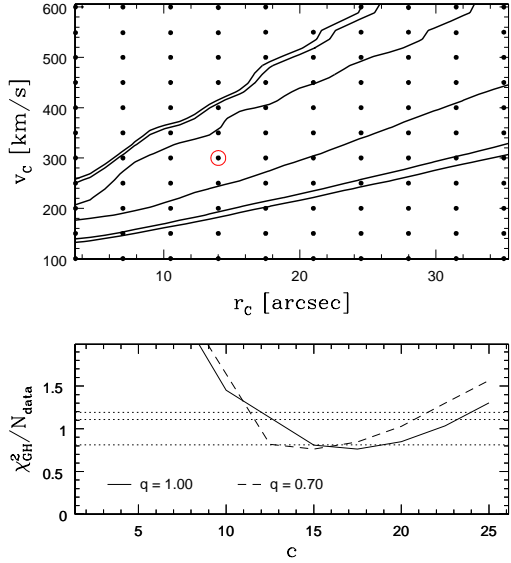}
\caption{As Fig.~\ref{nisconthalo}, but for libraries fitted to the kinematics of
NGC 4807. The best-fitting LOG-model is designated by the ring. In the lower panel results
for spherical (solid line) as well as flattened NFW halos are illustrated 
(dashed line: $q=0.7$).}
\label{niscont}
\end{figure}

In accordance with the isotropic rotator simulations the halo parameters $r_c$ and $v_c$ 
are not well constrained -- demonstrated by the $68$, $90$ and
$95$ per cent confidence contours in Fig.~\ref{niscont}. The tilted contour cones
are narrower than expected from Fig.~\ref{nisconthalo}. Note
however, that we have rescaled the $\chi^2_\mathrm{GH}$ and hence effectively reduced
the error bars as compared with the simulations. Therefore, also the shaded areas in 
Fig.~\ref{massstruc} are smaller than the corresponding in Fig.~\ref{halorecon}.

\begin{figure}
\includegraphics[width=84mm,angle=0]{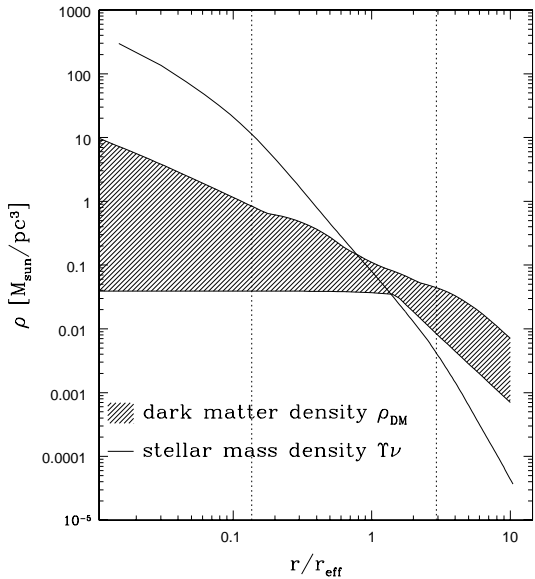}
\caption{Major-axis stellar mass-density according to the edge-on deprojection (solid line,
$\Upsilon = 3.0$) and data allowed dark matter densities. Both, LOG and NFW-halos are included
in the plot. (The shaded area is constructed analogously to Fig.~\ref{halorecon}.)}
\label{rhoconf}
\end{figure}

In the lower panel of Fig.~\ref{niscont} the goodness of fit 
$\chi^2_\mathrm{GH}/N_\mathrm{data}$ is plotted versus concentration $c$ for the one 
parameter family of NFW-profiles.
As in the simulations of Sec.~\ref{halop}, the halo flattening is not
constrained and with a best-fit $\chi^2_\mathrm{GH}/N_\mathrm{data} \approx 0.77$
spherical as well as flattened NFW-halos provide almost indistinguishable good fits to the 
data as the best-fitting LOG-halos. In the innermost regions of those libraries that match 
the observations, we find $\rho/\rho_\mathrm{DM} > 10$; the logarithmic density slope of the
best-fitting spherical NFW model is $\gamma=-1.8$ at the outermost major-axis data point and
$\gamma=-2.41$ at the outer edge of the orbit library. Hence, as already pointed out in
Sec.~\ref{halop}, the mass distributions of the data-allowed NFW models -- over the 
modelled spatial region -- resemble halos of LOG type. Fig.~\ref{rhoconf} illustrates the 
data-allowed range of dark matter densities along the major axis.
Note that the lower limit on $\rho_\mathrm{DM}$ is likely due to the limited range of 
profile shapes probed in our study. Dark matter distributions with negative central density 
slopes $\gamma<0$ (not tested here) might also provide acceptable orbit models.

\begin{figure}
\includegraphics[width=84mm,angle=0]{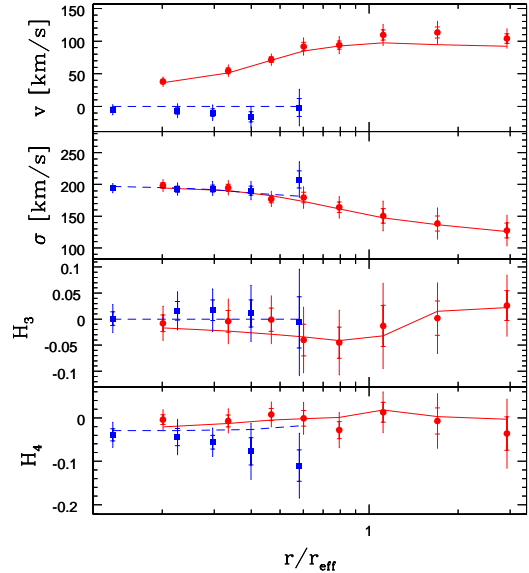}
\caption{As Fig.~\ref{ghfitbest}, but for the best-fitting model at
$i=50\degr$.}
\label{ghfiti50}
\end{figure}

Models at $i=50\degr$ are ruled out by more than $68$, but less than $90$ 
per cent confidence. The best-fitting
case is confronted with the kinematic observations in Fig.~\ref{ghfiti50}. Representative for
all low-inclination models, it fails to reproduce the minor-axis 
$H_4$-profile, but the mass structure of this model joins to the profiles in 
Fig.~\ref{massstruc} and the inferred structural properties of the galaxy are robust
against the assumed inclination.

\section[]{Stellar motions in NGC 4807}
\label{kin5975}

Visual inspection of the surfaces of section reveals that none of the orbit libraries 
for NGC 4807 contains a significant fraction of chaotic orbits. In the best-fitting
models, at all inclinations, no indication for chaos is detectable at all.

\begin{figure}
\includegraphics[width=84mm,angle=0]{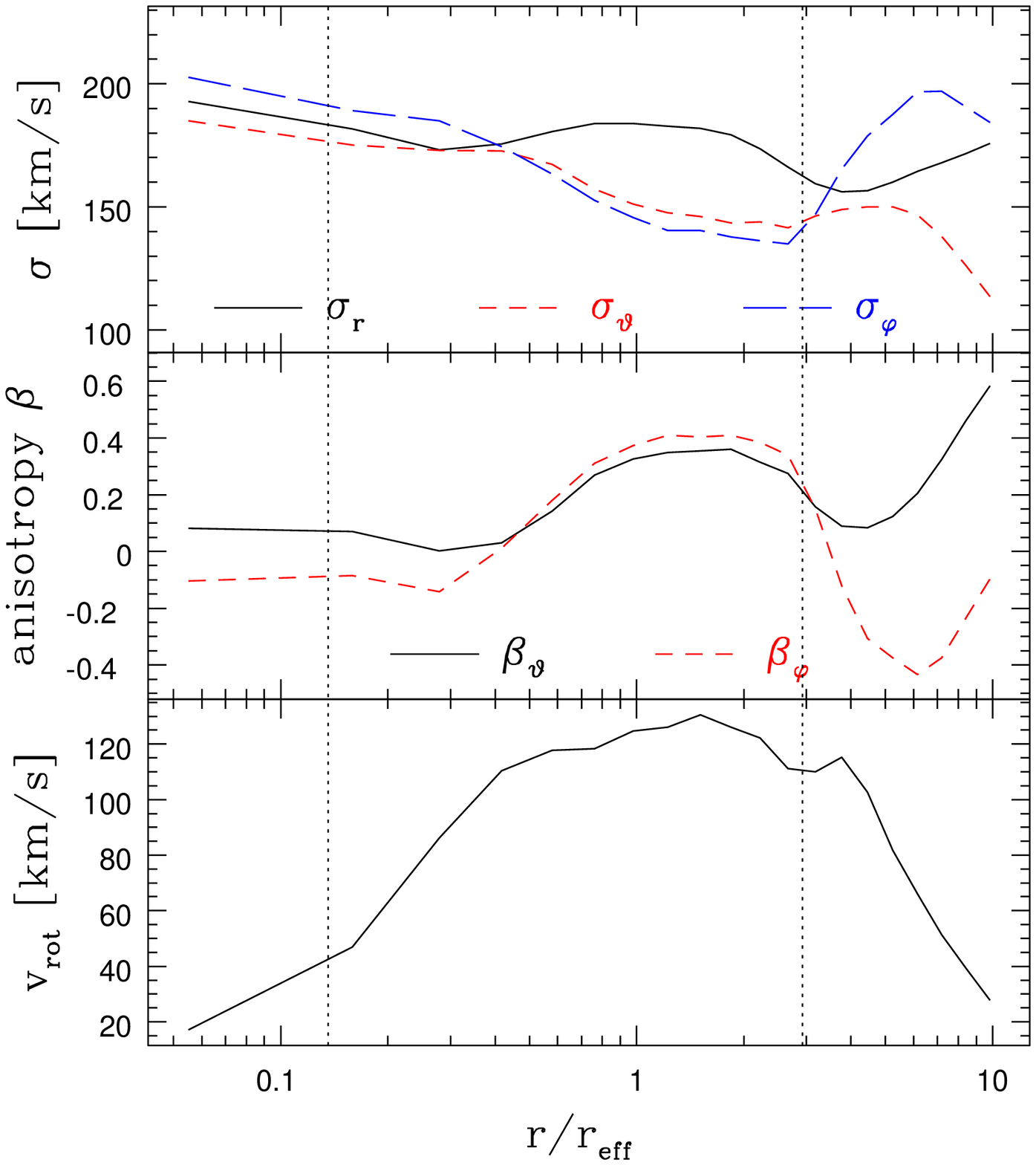}
\includegraphics[width=84mm,angle=0]{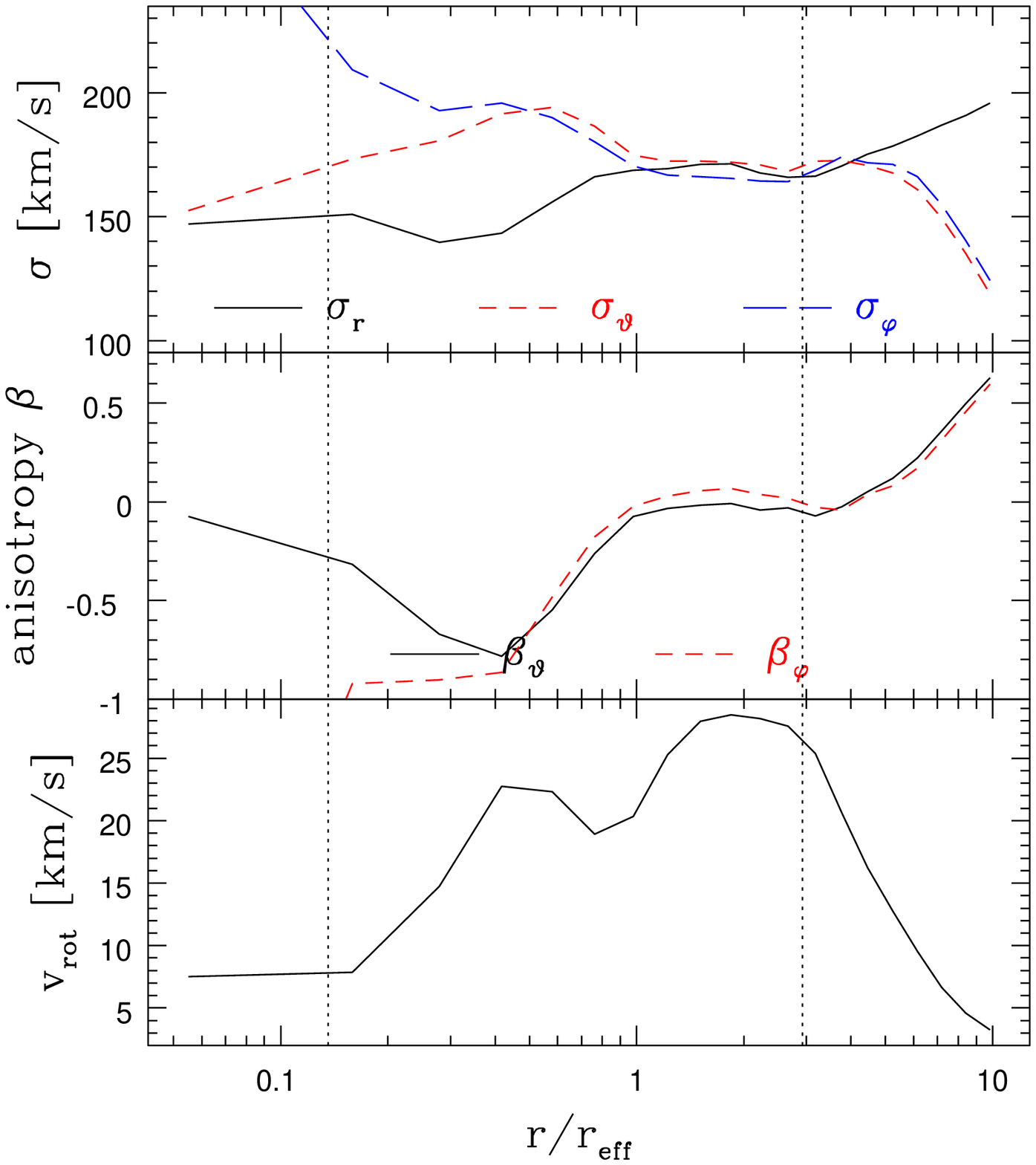}
\caption{Internal kinematics of the best-fitting halo model 
near the major axis (top panel; position angle $\vartheta=2.9\degr$) and
near the minor axis (bottom panel; $\vartheta=77.1\degr$), respectively. From top to bottom:
velocity dispersions $\sigma_r$ (solid), $\sigma_\vartheta$ (short-dashed) and
$\sigma_\varphi$ (long-dashed); velocity anisotropy $\beta_\vartheta$, $\beta_\varphi$; 
internal rotational velocity $v_\mathrm{rot}=v_\varphi$. Vertical, dotted lines 
indicate the innermost and outermost radius of kinematic data. Note, that the minor axis
data reach only out to $0.6 \, r_\mathrm{eff}$.}
\label{intdyn}
\end{figure}

The internal orbital structure of the best-fitting edge-on halo model 
near the major and minor axes is shown in Fig.~\ref{intdyn}. 
Around the equatorial plane, at radii $1 \, r_\mathrm{eff} < r < 3 \, r_\mathrm{eff}$, 
the model is characterized by radial anisotropy. Enhanced radial velocity dispersion 
$\sigma_r$ causes $\beta_\vartheta \approx \beta_\varphi \approx 0.3$, whereas
$\sigma_\vartheta \approx \sigma_\varphi$. Taking the large rotation velocity into account, 
energy in azimuthal motion turns out to be roughly equal to the 
energy in radial motion,
$\langle v^2_\varphi \rangle \equiv v_\mathrm{rot}^2 + \sigma_\varphi^2 \approx \sigma^2_r$.
On the other hand, motion perpendicular to the equator is suppressed. Approaching 
the central parts the velocity structure changes to isotropy and inside
$r<0.5\,r_\mathrm{eff}$ stays isotropic.

Close to the minor axis, NGC 4807 appears nearly isotropic in the outer parts 
$1 \, r_\mathrm{eff} < r < 3 \, r_\mathrm{eff}$ with decreasing radial dispersion 
towards the center, such that $\beta_\vartheta \approx \beta_\varphi < -0.5$ inside 
$r<0.5 \, r_\mathrm{eff}$.

\begin{figure}
\includegraphics[width=84mm,angle=0]{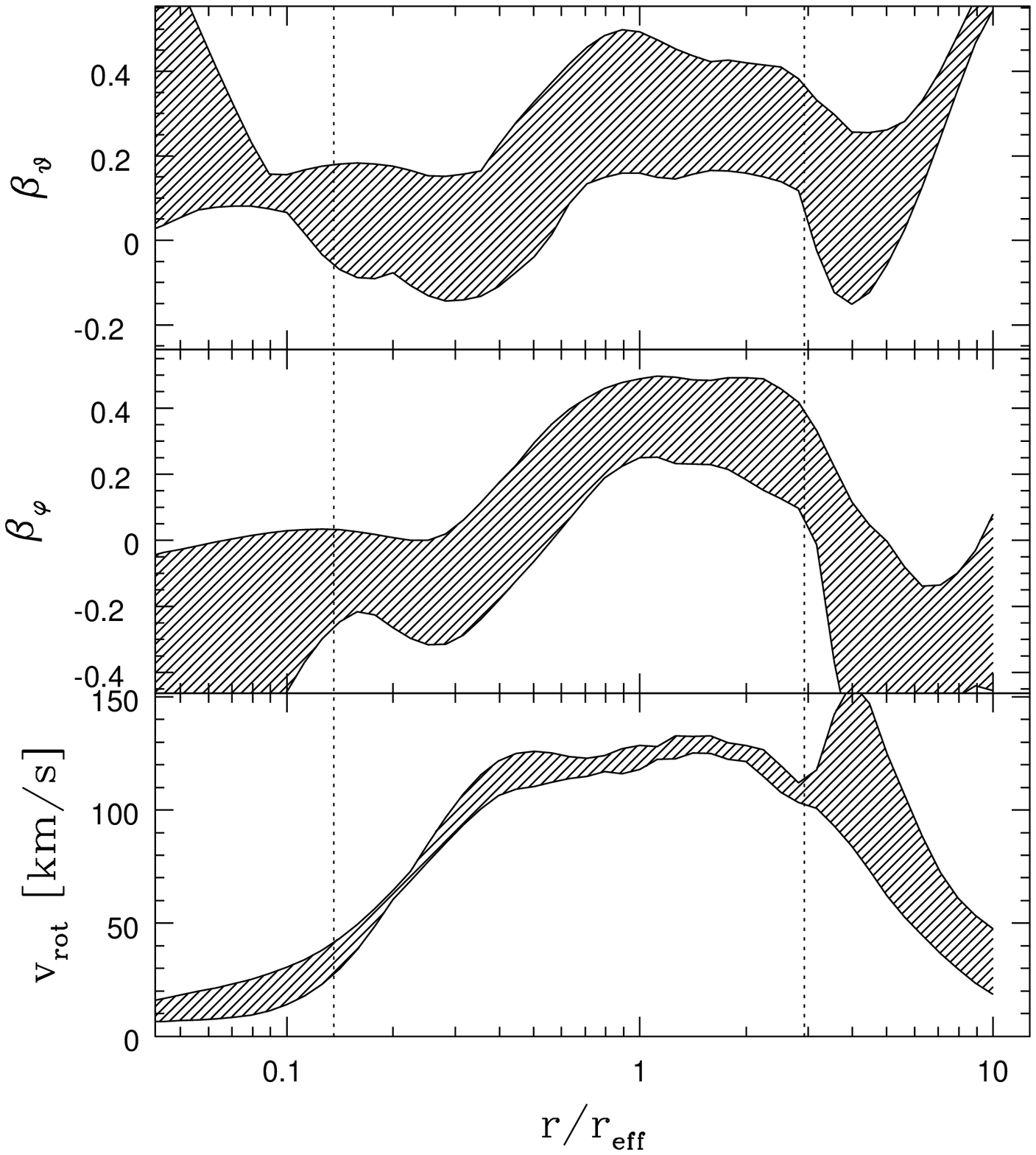}
\includegraphics[width=84mm,angle=0]{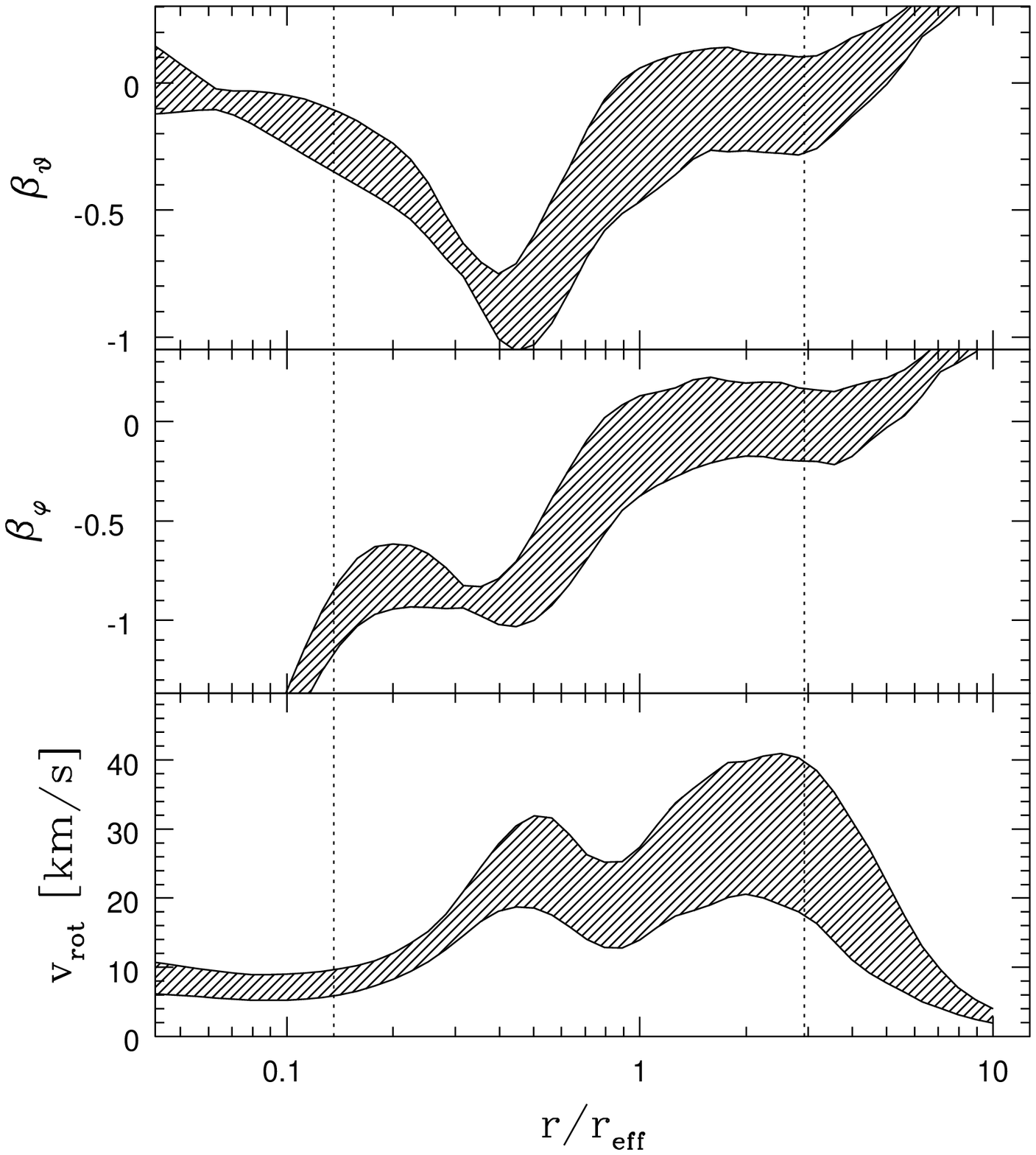}
\caption{Confidence regions for meridional velocity
anisotropy $\beta_\vartheta$, azimuthal velocity anisotropy $\beta_\varphi$ and
internal rotational velocity $v_\mathrm{rot}$ (upper panel: near major axis, position angle
$\vartheta = 2.9\degr$; lower panel: near minor axis, position angle $\vartheta=77.1\degr$) 
of NGC 4807. Shaded areas are constructed as for Fig.~\ref{halorecon}.}
\label{intdynconf}
\end{figure}

$68$ per cent confidence intervals analogous to Fig.~\ref{halorecon} for 
velocity anisotropies $\beta_\vartheta$ and $\beta_\varphi$ as well as internal rotation 
velocities near the major and minor axes are shown in Fig.~\ref{intdynconf}. 
Most tightly constrained is the major-axis rotation, rising from the central 
parts outwards until settling constant between 
$r_\mathrm{eff}$ ($v_\mathrm{rot} = 124 \pm 6$ km/s) and 
$3 \, r_\mathrm{eff}$ ($v_\mathrm{rot} = 107 \pm 4$ km/s). Velocity anisotropy is 
determined to at least $\Delta \beta \le 0.2$ and the trends in the anisotropy structure
of the best-fitting model in Fig.~\ref{intdyn} are clearly seen in all allowed orbit models.

\begin{figure}
\includegraphics[width=84mm,angle=0]{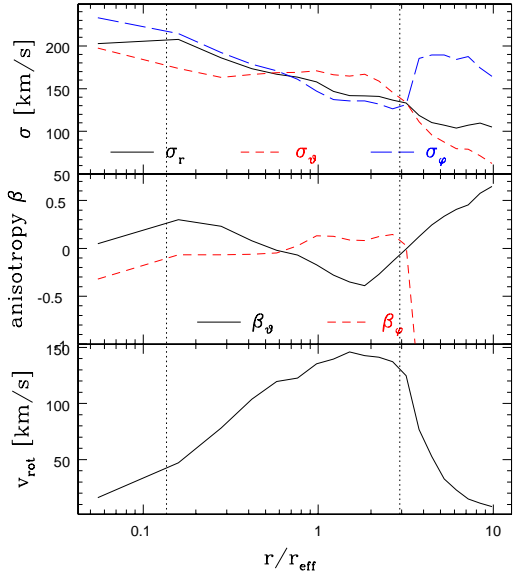}
\caption{Internal close-to-major axis kinematics (position angle $\vartheta=2.9\degr$) 
of the best-fitting orbit model at $i=50\degr$. Line definitions as in
Fig.~\ref{intdyn}.}
\label{intdyni50}
\end{figure}

Fig.~\ref{intdyni50} illustrates the internal close-to-major axis kinematics of the
best-fitting orbit model at $i=50\degr$. In contrast to the edge-on models 
$\beta_\vartheta \approx 0$ over almost the whole region constrained by observations, with 
a slight raise towards the center. Isotropy in the meridional plane 
($\sigma_r \approx \sigma_\vartheta$) is accompanied by suppressed azimuthal dispersion 
$\sigma_\varphi$. But in contrast to the edge-on model
of Fig.~\ref{intdyn}, the rotation of the actual model causes 
$\langle v^2_\varphi \rangle / \sigma^2_r \approx 1.7$, so that
this model is dominated by azimuthal motion near the equatorial plane.

\section[]{Phase-space structure of NGC 4807}
\label{phasstruc}

\begin{figure}
\includegraphics[width=84mm,angle=0]{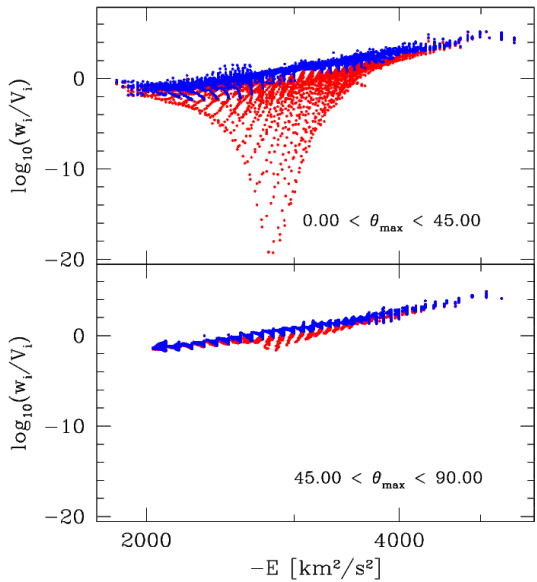}
\caption{Reconstructed distribution function of the best-fitting orbit model
at $\alpha=0.02$. Each dot represents the phase-space density along a single
orbit. Upper row: orbits confined to a konus with opening-angle 
$\vartheta_\mathrm{max} < 45\degr$ around the equator; lower row: remaining orbits.}
\label{dfbest}
\end{figure}

In Fig.~\ref{dfbest} phase-space densities $w_i/V_i$ 
of individual orbits are plotted against
orbital energy $E$ for the best-fitting orbit model of NGC 4807. The top panel only includes
orbits whose maximally reached latitude, $\vartheta_\mathrm{max}$,
is lower than $\vartheta_\mathrm{max}<45 \degr$. These orbits are confined to a cone with 
opening angle $\vartheta < 45\degr$ around the equatorial plane.
The remaining orbits are plotted separately in the bottom panel. As in 
Figs.~\ref{overfitting1} and \ref{overfitting2} the phase-space densities are scaled 
according to $\sum w_i = \sum V_i = 1$, where $w_i$ and $V_i$ are the orbital weights and
orbital phase volumes, respectively. 
Whereas the orbits maintaining the bulge of the galaxy follow a DF similar to the one 
shown in Fig.~\ref{overfitting1}, some of the orbits confined around the equatorial plane 
are strongly depopulated. Around $E \approx 3000 \,\mathrm{km}^2/\mathrm{s}^2$ phase-space
densities are up to 20 orders of magnitude below the main stream DF.

\begin{figure}
\includegraphics[width=84mm,angle=0]{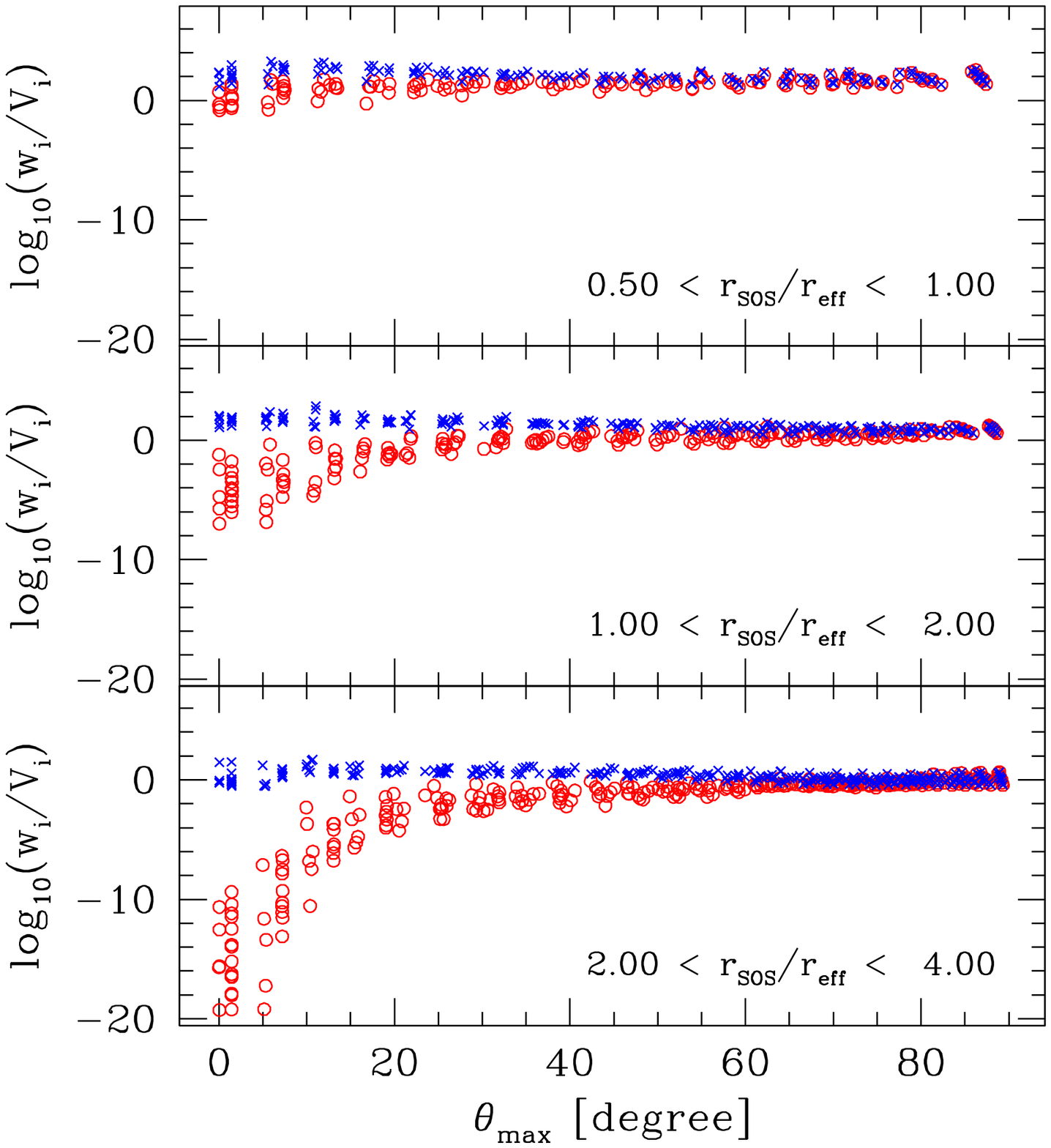}
\includegraphics[width=84mm,angle=0]{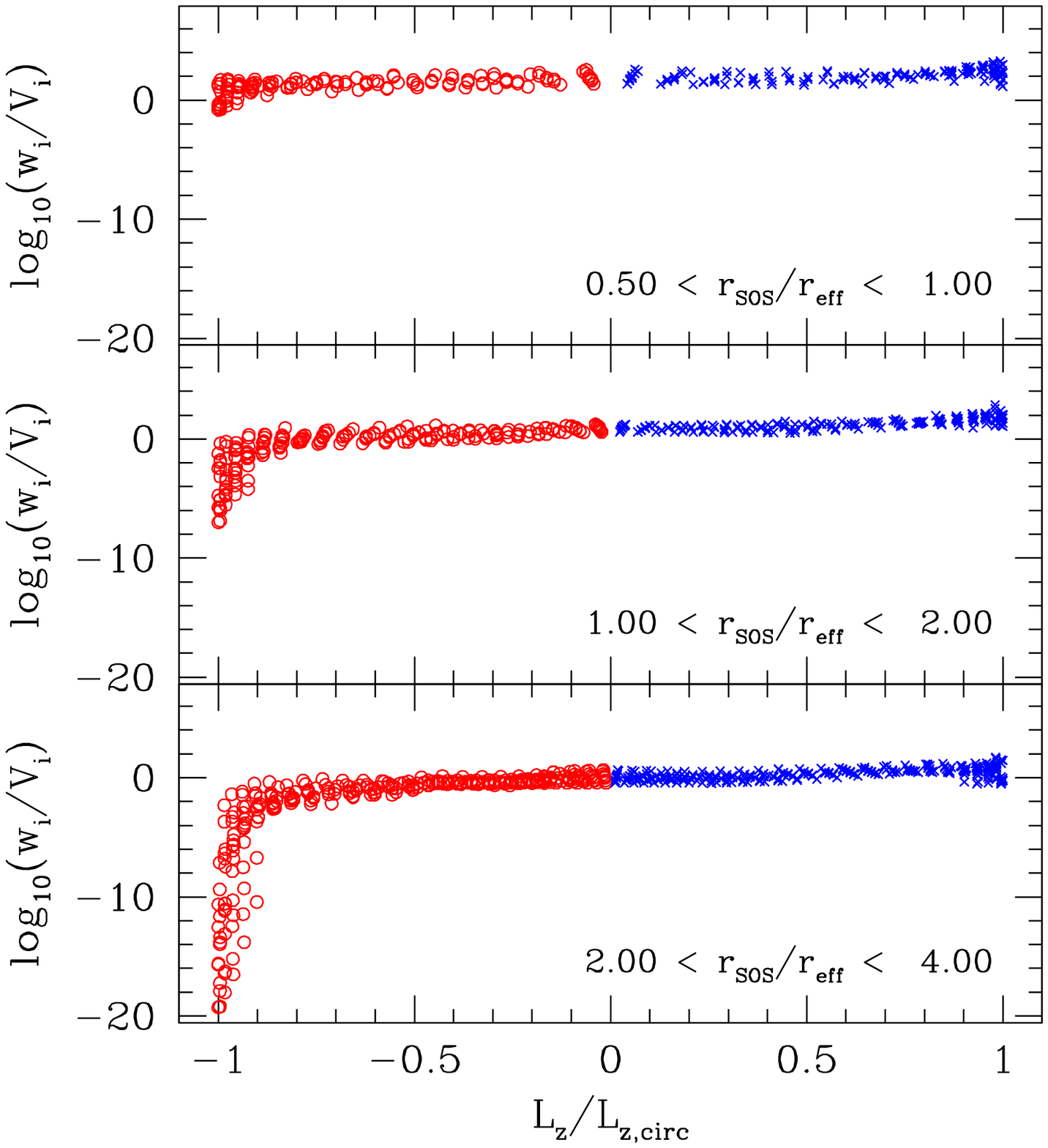}
\caption{The same model as in Fig.~\ref{dfbest}. 
Top panel: phase densities versus maximum orbital elongation 
$\vartheta_\mathrm{max}$ in the meridional plane. 
Bottom panel: phase densities versus ``circularity'' ($L_z$ scaled by the
angular momentum $L_{z,\mathrm{circ}}$ of a circular orbit with same 
energy). Crosses (open circles) denote orbits with positive (negative) $L_z$.
Orbits are extracted from different spatial regions of the model as designated in the
panels (see text for details). For each spatial region the number of prograde orbits
(crosses) equals the number of retrograde orbits (open circles).}
\label{dfreg}
\end{figure}

In the upper panel of Fig.~\ref{dfreg} orbital phase densities are plotted 
against maximum orbital
elongation $\vartheta_\mathrm{max}$ in the meridional plane. The three rows show orbits
located in three different spatial regions separately. The top row includes only orbits whose
invariant curves in the surface of section are confined to 
$0.5 \, r_\mathrm{eff} < r_\mathrm{SOS} < 1 \, r_\mathrm{eff}$\footnote{Here we use
$r_\mathrm{SOS}$ as a short cut for the radii of orbital equatorial
crossings, defining the surface of section. Fig. 1 in \citet{comadyn1} gives an example
of a typical surface of section in the actual context. The equatorial crossings
locate orbits in the meridional plane fairly well.}. The spatial
regions for the other two rows are 
$1 \, r_\mathrm{eff} < r_\mathrm{SOS} < 2 \, r_\mathrm{eff}$ and
$2 \, r_\mathrm{eff} < r_\mathrm{SOS} < 4 \, r_\mathrm{eff}$, respectively.
Crosses denote orbits with
positive $L_z > 0$ while open circles denote orbits with negative $L_z < 0$. From the
center outwards, orbits with negative angular momentum are progressively depopulated
as compared with their counterparts at positive $L_z$. The difference between prograde and
retrograde orbits is most noticeable near the equatorial plane and disappears completely 
for bulge orbits reaching up to higher latitudes. In the bottom panel of Fig.~\ref{dfreg} 
orbital phase densities are plotted against $L_z/L_{z,\mathrm{circ}}$, 
$L_{z,\mathrm{circ}}$ being the angular momentum of a circular orbit with the actual
energy. Phase-space densities are extracted in the same spatial regions as for the upper 
panel of the Figure. It is now apparent that the keel in the DF of Fig.~\ref{dfbest} is
caused by a depopulation of 
(retrograde) circular orbits (at $|L_z| \approx L_{z,\mathrm{circ}}$).

In fact, depopulation of retrograde near-circular orbits leading to the just described keel
appears in all data-allowed DFs, with a slight tendency to be less pronounced in flattened
halos. Even the DFs obtained assuming NGC 4807 is
inclined by $i=50\degr$ are featured by this structure, albeit there, the keel is not
exclusively made of retrograde orbits: also some of the prograde orbits are suppressed.
In the low inclination DFs a third prominent substructure in phase-space emerges
in form of a significant fraction of high-energy orbits (around 
$E \approx - 2500 \, \mathrm{km^2/s^2}$)
being suppressed by about 6 orders of magnitude in density, independent from $L_z$. This
structure partly overlays the keel and is 
caused by the distorted luminosity density at $i=50\degr$ since it shows up even in the
maximum entropy models ($\alpha \approx 0$) which are only forced to reproduce the 
density profile, but not fitted to the kinematics.

Orbits in the keel together with their prograde counterparts (approximately all
orbits $i$ with $w^{+}_i/w^{-}_i>99$, where $w^{\pm}_i \equiv w(E_i,\pm L_{z,i},I_{3,i})$,
$1\le i \le N_\mathrm{pro}$ and $N_\mathrm{pro}$ prograde orbits in the model)
make only $4.3$ per cent of the total model. Without these orbits, a slowly rotating bulge
($v \approx 70 \, \mathrm{km/s}$ at $r>r_\mathrm{eff}$) with a slightly peaked
velocity distribution ($H_4 > 0.04$ at $r>r_\mathrm{eff}$) appears 
(see also Sec.~\ref{dfdisc}). On the other hand, the extracted orbits rotate with about 
$200 \, \mathrm{km/s}$ and produce a narrow, low-dispersion LOSVD.
They develop a dumbbell-like structure extending to $2 \, r_\mathrm{eff}$ above the
equatorial plane. The fraction of extracted orbits additionally obeying 
$\vartheta_\mathrm{max} < 10\degr$ shrinks to $0.7$ per cent of the total model.


\section[]{Summary and Discussion}
\label{discussion}

\subsection{Regularized orbit models}
We have investigated how closely the internal mass distribution and kinematic structure
of a galaxy can be recovered from a sparse kinematic data set typical for our project
aimed at investigating a sample of flattened early-type galaxies in the Coma cluster.
The degree to which orbit models follow internal galaxy properties depends
on the amount of regularization applied in the fits. In the maximum entropy technique of
\citet{maxspaper} used here, a regularization parameter $\alpha$ controls the relative 
importance of entropy maximization (regularization) on the one side, and fit to the data
on the other. To find out which choice of $\alpha$ is optimal with respect to our goal of 
determining galaxy structural parameters, we simulated observationally motivated isotropic 
rotator models under realistic observational conditions. The models are based on a 
prototypical elliptical in our sample, NGC 4807. By varying $\alpha$ in the fits of 
appropriate orbit libraries to pseudo data of the reference models, the match between internal
velocity moments of input model and orbital reconstruction can be evaluated as a function
of regularization.

Our simulations 
indicate that the mass-structure of an elliptical can be recovered to about 15 per cent 
accuracy in terms of mass-to-light ratio and circular velocity curve, if a regularization 
parameter $\alpha=0.02$ is applied. For the dynamical models tested here, the optimal choice 
of $\alpha$ turns out to be roughly independent from the galaxy's gravitational potential.
The same accuracy as for the mass structure is achieved in the reconstruction of internal
velocity moments. On the other hand, halo flattening and 
galaxy inclination are only weakly constrained by our data.

Regularization biases orbit models towards some given idealized galaxy model, assumed
to represent the object under study reasonably well. For early-type
galaxies, regularization has been commonly implemented to isotropize the final fit. How
much of regularization is optimal in the reconstruction of a given galaxy depends
on the specific data set (spatial coverage, quality) on the one hand and the degree to
which the regularization bias approximates the given galaxy on the other. Isotropizing
the DF might, for example, be a good recipe for relaxed early-type galaxies but less
favorable for lenticulars with a significant, dynamically cold, component. In that
case either the value of $\alpha$ or, preferentially, the functional form of $S$ has to be
reconsidered. In any case, and for any regularization technique, 
the optimal balance between fit to data and smoothing of the DF can be examined
case-by-case from simulations similar to those described here.

\citet{Cre99} and
\citet{Verde02} have determined optimal regularization for an implementation 
of Schwarzschild's method with two-integral components by reconstructing similar DFs as 
used here. The resulting regularization has proven to yield plausible results in subsequent 
applications (e.g. \citealt{Cap02}; \citealt{Ver02}; \citealt{Cop04}; \citealt{Cre04}; 
\citealt{Kra05}). It is difficult to quantitatively compare the amount of regularization 
applied in these works with the results of our simulations since the regularization 
techniques differ. However, in each case the achieved accuracy in the 
recovery of test model parameters and the derived dynamical structure of real 
galaxies indicate that the applied amount of regularization is comparable.

\subsection{Luminous and dark matter in NGC 4807}
We then applied our code with the simulation-derived regularization to the elliptical
NGC 4807. The dynamical models require substantial dark matter in the outer parts
of the galaxy.
Evidence for dark matter in the form of flat circular velocity curves from 
(integrated) stellar kinematics of ellipticals has also 
been found by \citet{R97}, \citet{G98}, \citet{Ems99}, \citet{Cre00}, \citet{Kr00}, 
\citet{Sag00} and \citet{G01}. Dynamical models of five galaxies with planetary nebulae 
kinematics complementing stellar kinematics out to $\approx 5 \, r_\mathrm{eff}$ 
lead the authors to different appraisals: \citet{R03} argue for non-dark matter
models consistent with three galaxies observed by the planetary nebulae spectrograph,
while \citet{Peng04} require dark matter in NGC 5128 and the models of NGC 1399
also point at a dark halo \citep{Sag00}. 

Spherical models for the 21 round ellipticals in the sample of 
\citet{Kr00} reveal dark matter fractions of $10 - 40$ per cent at $1 \, r_\mathrm{eff}$ 
and dark mass equals luminous mass at roughly $2-4 \, r_\mathrm{eff}$ \citep{G01}, 
both comparable to our results for NGC 4807. The halo core density 
$\rho_c$ of our best-fitting model is about 30 per cent below the predictions of 
their $\rho_c-L_B$ relation (taking $M_B = -20.76$), although still consistent with it.
Among the allowed LOG-halos however, the core densities vary by a factor of ten and
taking into account NFW-fits, even halos $90$ times denser then the best-fitting model
are consistent with the data (cf. Fig.~\ref{rhoconf}). 
From the concentration of our best-fitting (spherical) 
NFW-halo nevertheless a relatively low formation redshift $z_f \approx 2.5$ follows. 
Still, we need a larger number of flattened ellipticals modelled in sufficient generality 
to recover the detailed properties and physical origin of their mass distributions.

\begin{figure}
\includegraphics[width=84mm,angle=0]{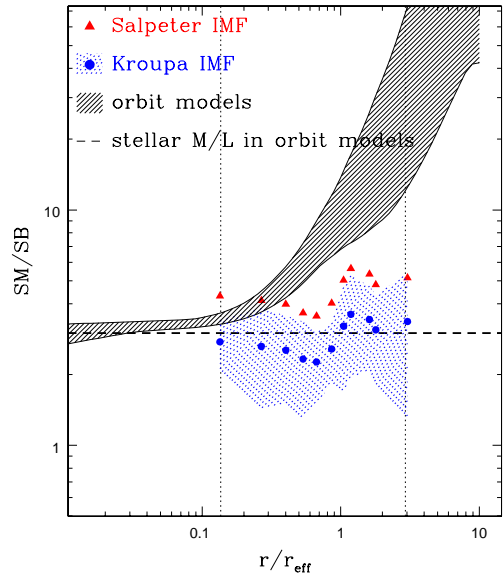}
\caption{Projected mass-to-light ratio along the major axis of NGC 4807.
Hatched area: 68 per cent confidence region of surface mass (SM) over surface brightness (SB)
for all orbit models with $\Upsilon = 3.0$ (dashed line). For comparison the projected stellar
mass-to-light ratios from population analysis (see text) are shown for two different IMFs: 
Salpeter-IMF (triangles), Kroupa-IMF (dots). The pointed area encompasses the statistically
allowed $\Upsilon$ based on the Kroupa-IMF. For the sake of clarity, the similar, but shifted,
error region for the Salpeter-based $\Upsilon$ is omitted.}
\label{projml}
\end{figure}

To crosscheck our mass decomposition for NGC 4807 
we compared stellar mass-to-light ratios of the orbit superpositions 
with stellar mass-to-light ratios determined completely independent. In 
Fig.~\ref{projml} mass-to-light ratios from stellar population models 
(\citealt{maraston98}; \citealt{daniel03}; \citealt{maraston04}) 
of NGC 4807's major-axis spectrum
(\citealt{coma1}; \citealt{Me03}) are compared to the best-fitting orbit models 
(dashed line: stellar mass-to-light ratio
in orbit model; hatched region: 68 per cent confidence region of surface mass SM over 
surface brightness SB in orbit models; different symbols refer to different 
initial-mass-functions (IMFs) underlying the populations). Systematic uncertainties
stemming from the unknown IMF are roughly comparable to the statistical errors, 
indicated for the Kroupa-IMF by the pointed area. Only libraries with
$\Upsilon = 3.0$ are taken into account in the figure. 

As the figure demonstrates, the stellar mass-to-light ratios determined dynamically
agree with stellar populations following a Kroupa IMF to $\Delta \Upsilon \approx 0.5$. 
This (1) confirms
our mass decomposition and (2) justifies a posteriori the assumption of a constant stellar 
mass-to-light ratio. The total surface mass of the dynamical models, as indicated by the 
shaded region in Fig.~\ref{projml}, exceeds the 
stellar contribution by far and underlines the evidence for dark matter in NGC 4807.
Since our best-fitting orbit models have LOG-halos (or NFW-halos imitating the mass 
distribution of LOG-halos over the kinematically sampled spatial region) 
they have maximum stellar mass (e.g. \citealt{G98}).
This is in accordance with most previous studies that did not require steep central
dark matter profiles (\citealt{G98}; \citealt{Ems99}; \citealt{Cre00}; 
\citealt{Kr00}; \citealt{Sag00} and \citealt{G01}). On the other hand, \citet{R97} found
such profiles consistent with NGC 2434.

\subsection{Comparing the kinematics: $\chi^2_\mathrm{GH}$ versus $\chi^2_\mathrm{LOSVD}$}
\label{chidisc}
The confidence intervals for the structural properties of NGC 4807 have been
derived from $\chi^2_\mathrm{GH}$ as defined in equation (\ref{chigheq}). However,
our models do not explicitly minimize $\chi^2_\mathrm{GH}$ but instead 
$\chi^2_\mathrm{LOSVD}$ as given by equation (\ref{chilosvd}).
In order to investigate how the properties of the best-fitting orbit model and
the corresponding confidence limits of the halo parameters depend on the
choice of $\chi^2$ we have analyzed the Monte-Carlo simulations as well as the models
for NGC 4807 also in terms of
\begin{equation}
\hat{\chi}^2_\mathrm{LOSVD} \equiv \chi^2_\mathrm{LOSVD}/\hat{N}_\mathrm{data},
\end{equation}
where $\hat{N}_\mathrm{data}$ is the number of velocity bins for which 
${\cal L}^{jk}_\mathrm{dat} > 0$ (cf. equation \ref{chilosvd}).

\begin{figure}
\includegraphics[width=84mm,angle=0]{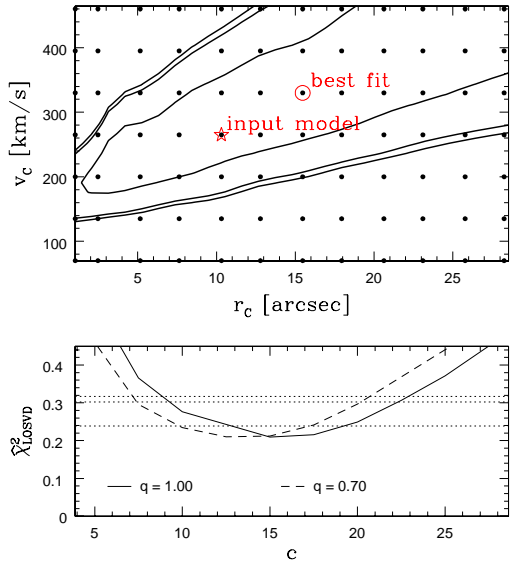}
\caption{As Fig.~\ref{nisconthalo}, but the confidence regions are derived from
$\hat{\chi}^2_\mathrm{LOSVD}$ (see Sec.~\ref{chivschi}).}
\label{nisconthalolosvd}
\end{figure}

Fig.~\ref{nisconthalolosvd} shows the best-fitting halo parameters and corresponding 
confidence intervals for the simulated isotropic rotator model of Sec.~\ref{halop}. 
The figure is as Fig.~\ref{nisconthalo}, besides that all confidence regions and 
model-with-data comparisons are computed in terms of $\hat{\chi}^2_\mathrm{LOSVD}$. Comparing
Figs.~\ref{nisconthalo} and \ref{nisconthalolosvd}, it turns out that both 
$\chi^2$-definitions yield identical best-fitting models (designated by 
the rings in the upper panels, and given by the minima of the two curves in the lower panels).
Concerning the confidence regions, however,
the computations based on $\hat{\chi}^2_\mathrm{LOSVD}$ lead 
to smaller confidence limits for both, the logarithmic and the NFW-halos.

\begin{figure}
\includegraphics[width=84mm,angle=0]{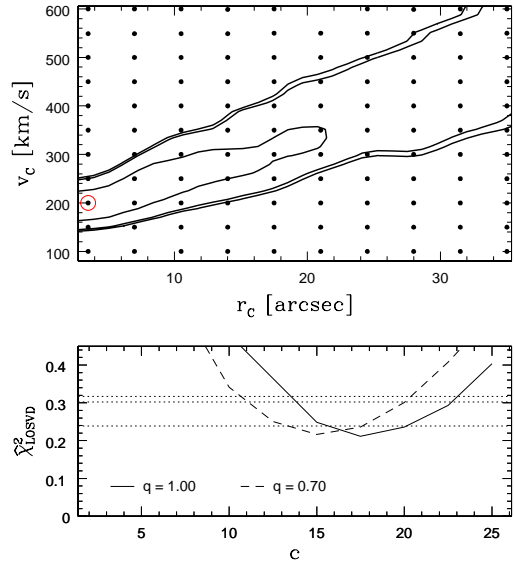}
\caption{As Fig.~\ref{niscont}, but the confidence regions are derived from
$\hat{\chi}^2_\mathrm{LOSVD}$ (see Sec.~\ref{chivschi}).}
\label{niscontlosvd}
\end{figure}

For NGC 4807, the analysis based on 
$\hat{\chi}^2_\mathrm{LOSVD}$ leads to the results shown 
in Fig.~\ref{niscontlosvd}. Again, besides the different $\chi^2$-definition applied, 
it is as Fig.~\ref{niscont}. The best-fitting NFW-halos
in Fig.~\ref{niscontlosvd} are the same as in Fig.~\ref{niscont}.
The shapes of the $\hat{\chi}^2_\mathrm{LOSVD}$ curves for
$q=1.0$ and $q=0.7$ in the bottom panel of Fig.~\ref{niscontlosvd} prefer halos slightly
more concentrated than in the $\chi^2_\mathrm{GH}$-case, but, as for the isotropic 
rotator simulations, the differences 
in the results obtained from $\chi^2_\mathrm{GH}$ and $\hat{\chi}^2_\mathrm{LOSVD}$ are small.

Regarding the best-fitting logarithmic-halo models (marked by the rings in the upper panels)
both $\chi^2$-calculations give consistent, but, in contrast to the Monte-Carlo simulations, 
not identical results. 
As already indicated in the NFW-fits, the logarithmic halo that best fits in the
sense of $\hat{\chi}^2_\mathrm{LOSVD}$ is more concentrated: the dark matter fraction
inside $r_\mathrm{eff}$ is $M_\mathrm{DM}/M = 0.35$, compared with $M_\mathrm{DM}/M = 0.22$
for the best-fitting model in the upper panel of Fig.~\ref{niscont}. 
A more striking difference between the upper panels of Fig.~\ref{niscont} and 
\ref{niscontlosvd} is the closure of the
68 per cent $\hat{\chi}^2_\mathrm{LOSVD}$-confidence contour to the upper right.
The contour differences maybe related to the ignorance of $\gamma_0$ in equation
(\ref{chigheq}), since the mismatch between model and data intensities progressively
increases from the lower left to the upper right in the upper panel of 
Fig.~\ref{niscont}.

Nevertheless, the comparison of the results obtained from $\chi^2_\mathrm{GH}$ and
$\hat{\chi}^2_\mathrm{LOSVD}$ reveals that both methods -- within the errors --
give the same results. The confidence limits quoted in this paper, based on equation 
(\ref{chigheq}), are the more conservative choice, but maybe slightly too pessimistic.

\subsection{The outer parts of NGC 4807}
\label{dfdisc}
Based on a recent refinement \citep{comadyn1} 
of the Schwarzschild code of \citet{maxspaper} and 
\citet{Geb00}, we recovered a depopulation of retrograde, near-circular orbits
in the phase-space DF of NGC 4807, giving rise to a keel when plotting orbital
phase-space densities against orbital energy. The prograde counterparts of the depopulated 
orbits form a dumbbell-like structure extending about $\approx 2 \, r_\mathrm{eff}$ above the
equatorial plane. 

\begin{figure}
\includegraphics[width=84mm,angle=0]{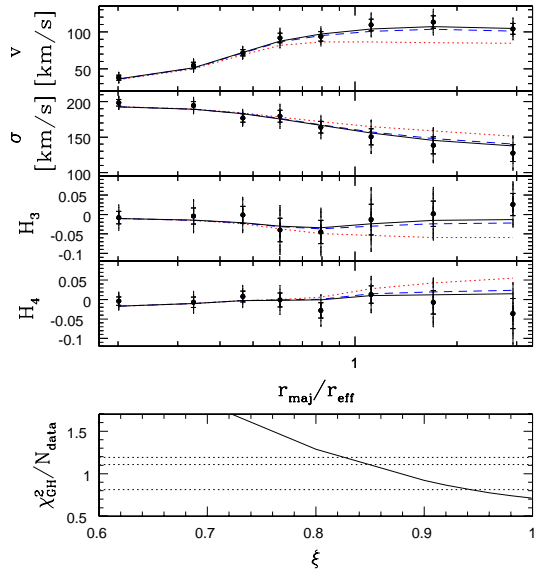}
\caption{Upper panel: projected major-axis kinematics of three distribution 
functions calculated in the gravitational potential of the best-fitting mass model: (1) all 
retrograde orbits in the keel depopulated completely (solid); relative fraction of 
light on prograde keel-orbit counterparts (2) $\xi=0.94$ (short-dashed) and (3) $\xi=0.7$ 
(dotted). Lower panel: goodness of fit 
(including the minor-axis LOSVDs, which are not affected by the redistribution of 
keel-involved orbits) as a function of the relative fraction $\xi$ of light on prograde 
keel-orbit correspondents; horizontal dotted lines indicate 68, 90 and 95 per cent confidence 
levels.}
\label{keelrem}
\end{figure}

To investigate what might cause the depopulation of retrograde orbits
in the outer parts of NGC 4807, we projected a sequence
of distribution functions in which all orbits involved in the keel ($w^{+}_i/w^{-}_i>99$, 
see Sec.~\ref{phasstruc}) are repopulated according to
\begin{equation}
\label{newweights}
\begin{array}{l}
w^{+}_{i,\mathrm{new}} \equiv \xi \times \left( w^{+}_i + w^{-}_i \right)\nonumber \\
w^{-}_{i,\mathrm{new}} \equiv (1-\xi) \times \left( w^{+}_i + w^{-}_i \right).
\end{array}
\end{equation}
Since $w^{+}_{i,\mathrm{new}} + w^{-}_{i,\mathrm{new}} = w^{+}_i + w^{-}_i$ this does
not alter the fit to the luminosity profile but just levels the relative fraction $\xi$ of 
light on the prograde and retrograde specimen of all keel-involved orbit pairs. Roughly 
speaking,
equation (\ref{newweights}) transforms the keel into a narrow cloud of points parallel
to the main-stream DF with an offset increasing with  $|\xi-0.5|$. Figure~\ref{keelrem}
shows the resulting projected (major-axis) kinematics for $\xi=0.7,0.94,1.0$. As expected,
shifting light from prograde to retrograde orbits reduces the amount of rotation $v$
and lowers $H_3$ in the outer parts, while at the same time leading to a larger velocity 
dispersion $\sigma$ and enhanced $H_4$. 
As shown by the goodness of fit in the lower panel of Fig.~\ref{keelrem}, the best fit to
the data is achieved for $\xi=1.0$ (all retrograde keel-orbits completely depopulated). 
Comparing the corresponding solid lines in Fig.~\ref{keelrem}
with the upper panel of Fig.~\ref{ghfitbest} reveals that $\xi=1.0$ provides essentially
the same fit as the best-fitting orbit model. Consequently, the retrograde keel-orbits 
can be regarded as completely depopulated in our models of NGC 4807.
Reducing the relative fraction of light on the prograde counterparts of keel-orbits to 
$\xi=0.94$ is only consistent with the observations at 
the 68 per cent confidence level. A further equalization to $\xi=0.7$ is 
already incompatible with the observed velocity profiles.

The outer major-axis LOSVDs of NGC 4807 might indicate weak triaxiality, for example in form 
of a weak, nearly end-on bar \citep[e.g.][]{Bur04}. A weak bar-like structure would also fit 
to the boxy appearance of the galaxy's outer isophotes. Assuming that NGC 4807 is slightly 
triaxial, the keel in the DF might be an artifact of the assumption of axisymmetry. Note, 
however, that NGC 4807 is consistent with being axisymmetric, albeit then the depopulation 
of single orbit families, especially around the equatorial plane, is hard to understand 
in the course of dynamical processes. 

Depopulation of retrograde orbits in the outer parts of the galaxy is accompanied by a change
in stellar ages from $\tau \approx 5 \, \mathrm{Gyr}$ inside $r<r_\mathrm{eff}$ to 
$\tau \approx 10 \, \mathrm{Gyr}$ at larger radii \citep{Me03}. 
Towards $3 \, r_\mathrm{eff}$ stellar
ages become uncertain and no clear trend is visible. It seems also possible, that
the keel in the DF is an artifact related to a distinct stellar component and an 
associated (slight) change in $\Upsilon$. It should be 
noted that because the orbit models are fitted to deprojections and stellar kinematics, 
the derived  DFs characterize the amount of {\it light} per volume element in phase-space, 
not the {\it mass}-density in phase-space. Accordingly, depressions in the DF might (at 
least partly) reflect enhanced $\Upsilon$.

A distinct stellar component in the outer parts must not necessarily take the form of a bar, 
but could also be a faint axisymmetric stellar disk. Indeed, the photometry as well as the 
orbit models suggest that NGC 4807 is close to edge-on, but a dynamically cold, outer stellar 
disk made of only $0.7$ per cent prograde orbits at latitudes $\vartheta < 10 \degr$ 
(see Sec.~\ref{phasstruc}) is consistent with the non-disky isophotes ($\max a_4 < 1$).
A disk carrying 2 per cent of the total luminosity would show up
with $\max a_4 \approx 4$ when seen exactly edge-on and
$\max a_4 < 1$ for such a disk would require $i<75\degr$ \citep{R90}. 
From our simulations it seems impossible to distinguish kinematically between $i=70\degr$ and
$i=90\degr$ and, consequently, there is some freedom for an outer disk in NGC 4807.

In any case further investigations are necessary to verify the significance of the 
keel-structure in the DF of NGC 4807.
(1) Kinematic measurements along intermediate position angles in the galaxy would likely 
put more constraints on the DF. (2) Detailed comparison of orbit superpositions with 
realistic galaxy models are required to estimate how relics of triaxiality, faint disks and 
multi-component (multi-$\Upsilon)$ structures show up in our axisymmetric models. 
These questions will be addressed in a forthcoming publication.

\subsection{Internal stellar kinematics of NGC 4807}
Along the major-axis, NGC 4807 is mildly radially anisotropic. Radial anisotropy has been
found in a number of ellipticals (\citealt{G98}; \citealt{Mat99}; \citealt{Cre00}; 
\citealt{Geb00}; \citealt{Sag00}). The amount of anisotropy ($\beta \approx 0.3$) in the 
outer parts as well as the isotropy of the central region (inside $0.3 \, r_\mathrm{eff}$)
matches with the typical anisotropy structure found for round, non-rotating ellipticals 
by \citet{G01}. Along the minor axis, NGC 4807 is dominated by tangential motions. 
\citet{Geb03} report tangential anisotropy for some galaxies (mostly enhanced azimuthal, 
but suppressed meridional dispersions) and also \citet{Dej96}, \citet{Cre98}, \citet{Sta99}, 
\citet{Cap02}, \citet{Ver02} and \citet{Cop04} find predominantly tangential motions.
For NGC 3115 \citet{Ems99} notice $\sigma_z > \sigma_r$. All these studies are based on
data sets with different spatial sampling and use various dynamical modelling techniques, 
particularly differing in the amount and functional form of the applied regularization. 
As for the mass structure and the dark matter properties, a large and homogeneous sample of 
galaxies is needed to address the physical processes shaping elliptical galaxies.

\section*{Acknowledgments}
We thank the referee Tim de Zeeuw for constructive comments that helped to improve the
manuscript. J. Thomas acknowledges financial support by the 
Sonderforschungsbereich 375 ``Astro-Teilchenphysik'' of the Deutsche Forschungsgemeinschaft.

\bsp

\label{lastpage}

\end{document}